\documentclass[a4paper,fleqn]{cas-dc}
\usepackage[noend]{algpseudocode}
 \usepackage{
  booktabs,
  color,
  float,
  svg,
  wrapfig,
  graphics,
  graphicx,
  subcaption,
  algpseudocode,
  tikz,
  xurl,
  url,
  enumitem,
  mfirstuc,
  colortbl
}

\setlist[itemize]{nosep}

\newcommand*\circled[1]{\tikz[baseline=(char.base)]{
            \node[shape=circle,draw,inner sep=1pt] (char) {#1};}}
            
\newcommand{\newtext}[1]{#1}
\newcommand{\E}[1]{\scalebox{0.6}{$\times$}\scalebox{0.8}{10}${}^{#1}$}

\begin{document}

\let\WriteBookmarks\relax
\def\floatpagepagefraction{1}
\def\textpagefraction{.001}

\shorttitle{Carbyne: An Ultra-Lightweight DoS-Resilient Mempool for Bitcoin}    

\shortauthors{HB Haq, ST Ali, A Salman, P McCorry, SF Shahandashti}  

\title [mode = title]{Carbyne: An Ultra-Lightweight DoS-Resilient Mempool for Bitcoin}  

\author[label1]{Hina Binte Haq} 
\author[label1]{Syed Taha Ali}
\author[label2]{Asad Salman}
\author[label3]{Patrick McCorry}
\author[label4]{Siamak F. Shahandashti}

\cormark[1]


\ead{siamak.shahandashti@york.ac.uk}

\credit{
    Hina Binte Haq: Conceptualization, Investigation, Methodology, Visualization, Writing - original draft; 
    Syed Taha Ali: Conceptualization, Supervision, Writing -- review \& editing; 
    Asad Salman: Data curation; 
    Patrick McCorry: Conceptualization; 
    Siamak F. Shahandashti: Validation, Writing -- review \& editing
}

\affiliation[label1]{organization={National University of Sciences and Technology (NUST)}, 
            city={Islamabad}, 
            country={Pakistan}}

\affiliation[label2]{organization={X (formerly Twitter)}, 
            country={USA}}

\affiliation[label3]{organization={Arbitrum}, 
            city={London}, 
            country={United Kingdom}}

\affiliation[label4]{organization={University of York}, 
            city={York}, 
            country={United Kingdom}}

\cortext[1]{Corresponding author}

\nonumnote{A preprint of this paper is available on arXiv.}

\begin{abstract}
The Bitcoin mempool plays an integral role in transaction processing and propagation through the network. Frequent transaction congestion events, as well as spam and dust attacks can clog the mempool, leading to dropped transactions, processing delays, and increased transaction fees. Moreover, increasing transaction loads on the network result in higher resource costs to operate full nodes, thereby restricting Bitcoin's network footprint and negatively impacting its overall health and performance. In this paper, we present Carbyne, a novel mempool optimization scheme, which uses counting bloom filter constructions to adapt to increased transaction flows, thereby making nodes resilient to congestion and spam and dust attacks. We implement Carbyne in C++ and benchmark its performance using a novel data set of Bitcoin mempool activity over a 90-day period. We dramatically reduced the mempool's memory consumption by up to two orders of magnitude (from 300 MB to 3 MB) while verifying and forwarding transactions with 99.9\% fidelity and a slight increase in computational load. We simulate extensive spam attacks on Carbyne and demonstrate that mempool loads of 1 GB can be accommodated in as little as 10 MB. Carbyne does not necessitate a hard fork, it will help deploy high-functioning nodes on resource-constrained platforms, and it may also be adapted to other cryptocurrencies.
\end{abstract}


\maketitle

\section{Introduction}
\label{sec:introduction}
Bitcoin and cryptocurrencies are achieving mainstream success. Bitcoin has a market capitalization of over \$806 billion, and the top 5 cryptocurrencies, collectively account for a valuation over \$ 1.4 trillion~\cite{coinmarketcap}. Bitcoin exceeded 130 million users in 2021 amid predictions it will grow to a billion users in the next four years~\cite{nasdaqprojection}. Some 46 million Americans, 17\% of the adult population, now own Bitcoins~\cite{nasdaqusers}, and 36\% of small-medium businesses accept it~\cite{hsb}.

This popularity has motivated considerable research on the scalability of Bitcoin, to improve Bitcoin's transaction throughput (e.g.~\cite{gilad2017algorand}~\cite{eltoo}) and reduce its growing memory requirements (e.g.~\cite{jiang2019bzip}~\cite{dryja2019utreexo}). Interestingly, these problems may also be interpreted as security challenges that typically manifest as degradation or denial of service attacks. However, this particular research domain has received little attention from a security perspective. Is it possible to reconcile these twin aims, scalability and security, to develop solutions to help Bitcoin grow in ways that also harden the Bitcoin network in terms of security and robustness?


In this paper, we consider the motivating example of growing transaction loads and network congestion. Bitcoin nodes use local memory (RAM) to store state information, primarily the unspent transaction output (UTXO) set and the Bitcoin transactions memory pool (mempool). The mempool plays an integral role in transaction processing: it logs unconfirmed transactions and assists in their propagation through the network. Storing these transactions in RAM as opposed to disk is efficient, as disk latency slows transaction processing by up to two orders of magnitude~\cite{jiang2019bzip}.  

The average number of transactions in the Bitcoin mempool, has more than doubled in 2021~\cite{mempool-count}. Transaction congestion frequently results in significant increase in local RAM utilization which may negatively impact the overall performance of the network. If the queue of unconfirmed transactions exceeds a node's allocated mempool size limit, the node starts to drop or ignore transactions. This in turn causes a degradation in service in terms of prolonged processing delays and increased transaction fees~\cite{nov17}~\cite{dec2017}~\cite{nov19}. For instance, a large spike in transaction volumes in October, 2020, led to some 145k transactions being stalled and median transaction fees reaching a three-year peak of \$11.66~\cite{coingate2020your}. There are frequent reports of mempool congestion on the Ethereum network, due in part to the popularity of decentralized finance applications~\cite{neal20ethereum}. Surge in transaction loads have led to extended network outages on Solana~\cite{binance}.

The mempool's susceptibility to congestion is also a highly potent attack vector which parties exploit by deliberately flooding the network with very low-cost transactions to clog the mempool. These activities, variously referred to as \emph{stress tests}, \emph{spam campaigns}, and \emph{dust attacks}, may be considered effective Denial of Service (DoS) attacks. One spam campaign in October, 2015 grew the mempool to nearly 1 GB, with over 88,000 transactions, and reportedly caused 10\% of nodes in the Bitcoin network to crash~\cite{baqer2016stressing}.

Moreover, sophisticated attacks have emerged which directly exploit mempool dynamics for profit. For instance, in March, 2020, when large sell-offs in crypto markets triggered collateral auctions on the Ethereum-based MakerDAO platform, unknown actors flooded the network with low-cost transactions~\cite{Dale2020Jul}. Many bidders could not adapt to the resulting spike in gas prices in time, thereby enabling the attackers to sweep the auctions with \$0 bids, and pocket \$8.3 million. In 2022, the Solana network was flooded by ``complex-compound-instruction transactions'', preventing certain users from updating collateral positions in time, suffering liquidation, and abandoning Solana in protest. \cite{haqshanas}

In response, researchers have devised mempool eviction strategies which screen incoming transactions for various features, including size, age, and fee, to identify and drop potential dust transactions~\cite{saad2020contra}~\cite{wang2018anti}. However, this approach has significant shortcomings: first, the solutions developed thus far have significant false positive rates, resulting in legitimate transactions being classified as spam, which some argue may constitute a DoS attack in its own right~\cite{baqer2016stressing}. Second, attackers can discover filter heuristics and potentially craft spam transactions to evade them.

We propose Carbyne, a novel and flexible Bitcoin mempool optimization scheme to provide resilience against escalating transaction loads and spam and dust attacks. Carbyne uses space-efficient probabilistic data structures to store fingerprints of unconfirmed transactions. This reduces the mempool to a fraction of its size in RAM, while still verifying and forwarding incoming transactions with high fidelity. 

Our key insight is that the two prime functions of the mempool, \emph{transaction forwarding} and \emph{transaction inventory}, can be dissociated, to prioritize one function over the other. A similar philosophy has commonly been employed to build lightweight clients, such as pruned nodes, simplified payment verification (SPV), EPBC \cite{xu2017epbc}, FlyClient \cite{bunz2020flyclient}, etc.

Moreover, Carbyne users can still choose to maintain a subset of transactions in RAM to mine blocks or to bootstrap new nodes on the network. Decoupling forwarding and inventory functions gives users the flexibility to control RAM usage and still contribute to the network in an accurate and efficient manner as per the resources available to them. 

Thus, Carbyne also helps address the related but overlooked problem concerning Bitcoin's growing local memory consumption. A sustained increase in transaction loads results in higher resource costs for users to operate Bitcoin nodes and, unlike miners, these parties are not incentivized for their contribution. This in turn restricts the footprint of the Bitcoin network, and may render it vulnerable to certain attacks. Our solution can help develop lightweight clients and alleviate these growing costs. Thus, Carbyne is not meant to supplant full nodes - but to lower entry costs for certain parties. Our specific contributions are: 

\begin{itemize}
\item 
\emph{We describe Carbyne, an optimized DoS-resilient mempool construction for Bitcoin} which uses multiple counting bloom filters to replicate the mempool's core inventory functions. Carbyne reduces the mempool space requirements in local memory by up to two orders of magnitude while still processing unconfirmed transactions with very high fidelity.
    
\item 
\emph{We implement Carbyne in C++ and benchmark its performance} on a novel dataset of Bitcoin mempool activity logged over a 90-day period, comprising approximately 29 million distinct transactions -- an independent contribution and useful for various applications apart from Carbyne (see Appendix~~\ref{sec:appendix_MempoolStateDataset}).  Researchers can use this dataset and the accompanying scripts to efficiently reconstruct Bitcoin mempool state. We define various metrics to undertake a fine-grained comparison of Carbyne versus the Bitcoin Core mempool. We also provide multiple parameters to fine-tune the trade-off between memory usage, fidelity, and robustness when deploying Carbyne.


\item 
\emph{We simulate extensive congestion and spam attacks} to demonstrate that Carbyne easily copes with high transactions loads, up to a threefold increase over the maximum flows ever witnessed in the Bitcoin network. We further propose mechanisms to dynamically adapt to rising transaction rates with modest computational and memory overheads.
\end{itemize}

We achieve a dramatic reduction in memory consumption: we are able to fit approximately 300\,MB worth of Bitcoin transactions in 3\,MB while processing them with 99.915\% accuracy. If 12 MB are allocated for Carbyne, accuracy increases to 99.997\%. In our simulated congestion experiments, 600,000 transactions, which would ordinarily require over 1\,GB, can be handled in under 10\,MB.

Carbyne has some limitations: it does not resolve block congestion issues or directly reduce transaction fees. Carbyne does not detect or evict spam but it can easily be integrated with spam filtering mechanisms ~\cite{saad2020contra}~\cite{wang2018anti}. The use of probabilistic data structures results in a minute false positive rate in the form of dropped transactions, which are significantly smaller than that reported in related work.

There are other considerable benefits: Carbyne can be deployed without necessitating a fork. Carbyne enables resource constrained systems (e.g. Raspberry Pi, smartphones ~\cite{htc}) to run high-performing functional mempools which validate and propagate incoming transactions, and can scale to handle congestion and spam attacks. Moreover, this application of bloom filters may hold value for other cryptocurrencies, such as Litecoin, Ethereum, Ripple, and Solana.

Our solution is orthogonal to other space-efficient optimizations for Bitcoin, including pruning, Segwit, and UTXO optimizations, and may therefore be combined with these strategies to aggregate their individual benefits. 

We believe we are the first to rethink the structure of the mempool to combat congestion and spam attacks. Carbyne aligns with other ongoing attempts to harden the overall security and resilience of the Bitcoin network (such as MIT's Bitcoin Software and Security Effort \cite{dci}). Our work also contributes to local memory optimization, which is a neglected domain in the research literature.

The paper is organized as follows: In \S\ref{sec:background}, we present essential background material. We describe Carbyne in \S\ref{sec:proposed} and present our dataset in \S\ref{sec:dataset}. This is followed by experimental results, discussion and security analysis in \S\ref{sec:results}. In \S\ref{sec:stresstest}, we simulate congestion and spam attacks. We conclude in \S\ref{sec:conclusion}. 

\section{Background}
\label{sec:background}
In this section we overview spam attacks, discuss prior work and present our threat model. We also describe the mempool and its structure. 

\subsection{Spam, Stress Tests, and Dust Attacks}
The first major campaign, advertised as a \emph{``stress test"}, occurred in July, 2015, with a daily average of 150,000 pending transactions. Another campaign in October, 2015 increased mempool size to nearly 1 GB, with more than 88,000 pending transactions. This incident reportedly knocked up to 10\% of Bitcoin nodes offline, many of which were running on memory-constrained platforms like Raspberry Pi~\cite{baqer2016stressing}. 

\begin{figure}
    \centering
    \begin{subfigure}{0.50\textwidth} 
        \includegraphics[width=8.5cm, height=3cm]{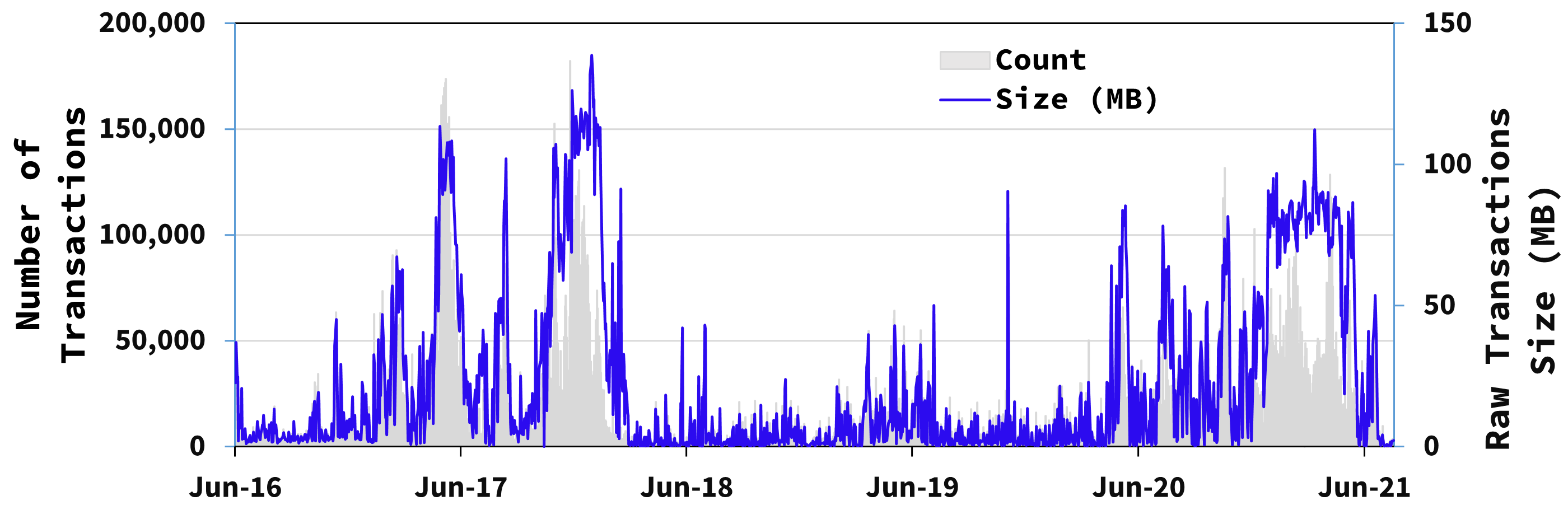}
            \vspace{-2mm}
        \caption{Mempool size and transaction count}
              \label{fig:alltime}
              \vspace{1mm}
   \end{subfigure}
    \hfill
    \begin{subfigure}{0.50\textwidth} 
        \includegraphics[width=8.5cm, height=2cm]{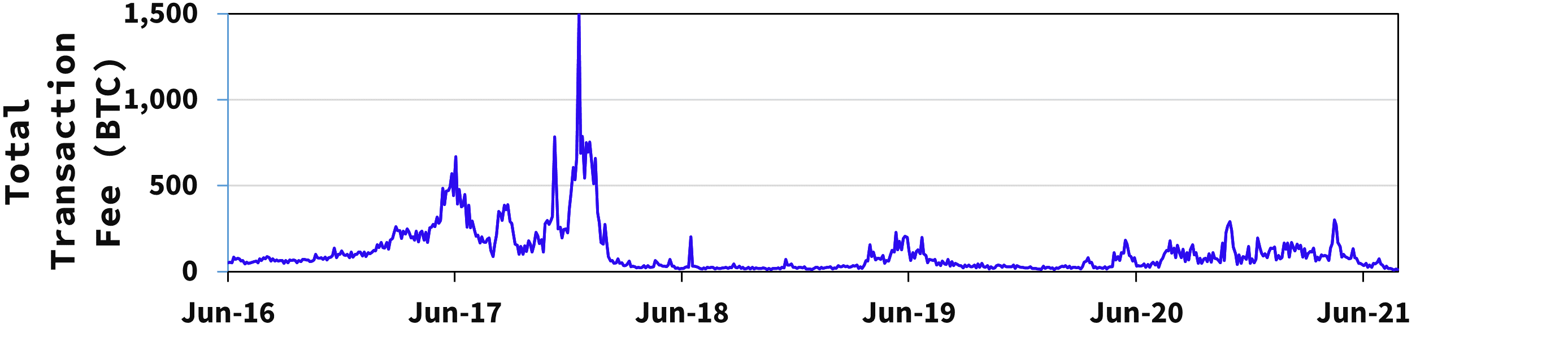}
        \vspace{-2mm}
        \caption{Transaction fees}
        \label{fig:txfeebtc}
    \end{subfigure}
    \label{fig:historical}
         \vspace{-4mm}
    \caption{Bitcoin mempool \& transaction fee trends~\cite{mempool-size}}
    \label{fig:mempool_tx_fee_trends}
\end{figure}

Numerous congestion and spam incidents have occurred since. In November, 2017, with the cancellation of the SegWit2x hard fork, many users started selling their coins, causing the mempool size to exceed 182,000 transactions and 125 MB in raw transaction size~\cite{nov17}. In December 2017, a DoS attack on Bitfinex inflated the mempool over 80,000 pending transactions and 135 MB in raw transaction size~\cite{dec2017}. In November 2019 the mempool spiked to over 115,000 pending transactions and 90 MB in raw transaction size when Binance moved its coins in bulk~\cite{nov19}. Fig.~\ref{fig:mempool_tx_fee_trends} shows these spikes in transaction rates and fees.

Other sources of spam include mixing services that use dust as a promotional tool~\cite{bestmixer}. Dust attacks have also been used as a tactic to compromise privacy~\cite{baqer2016stressing}. Games like Satoshi Dice also generate large numbers of low-value transactions that some categorize as `dust-like'~\cite{bruce2014mini}.

Other cryptocurrencies with a mempool-based architecture are also vulnerable to congestion and spam attacks. In 2019, over 200,000 Litecoin wallets received spam transactions as part of a publicity stunt by a mining pool~\cite{litecoin}. The attack on the Ethereum-based MakerDAO platform specifically focused on spiking transaction fees so that bidders using automated scripts could not adapt in time to compete in collateral auctions~\cite{Dale2020Jul}. In September 2021 the Solana network stalled for 17 hours as bots generated 400,000 transactions per second, overflowing the transaction processing queue, and causing some validator nodes to crash \cite{solana_twitter}. 

Our solution, in principle extends to other cryptocurrencies, including Litecoin, Ethereum, and Solana. We discuss these applications in \textbf{\S\ref{sec:othercrypto}.}

\subsection{Prior Work}
\label{sec:related}


Carbyne contributes to the literature on improving the security, resilience, and health of the Bitcoin network. It also aligns with research on scalability that improve Bitcoin's transaction throughput (e.g.~\cite{segwit}\cite{gilad2017algorand}\cite{wang2019sok}\cite{poon2016bitcoin}) and ongoing efforts to develop lightweight clients (e.g.~\cite{zamyatin2020txchain}~\cite{bunz2020flyclient}~\cite{kiayias2020non}\cite{cao2020cover}).



The mempool has been largely overlooked in the research literature. Baqer et al. conducted a post-attack analysis of one of the first spam attacks on Bitcoin in 2015~\cite{baqer2016stressing}. They use clustering to classify approximately 23\% of the observed transactions as spam. They explicitly identify the mempool as a vulnerable component and recommend evicting transactions to ``relieve the pressure''. The authors suggest Bitcoin  implements a \texttt{mintxfee} per output similar to Litecoin or a dynamic fees model to counter spam.

However, the authors also caution that filtering strategies can misclassify legitimate transactions as spam, and even a small 1--2\% false positive rate amounts to a DoS attack in its own right. Furthermore, attackers can gain knowledge of filtering parameters and craft transactions to evade them. 

Transaction eviction is the primary theme in papers that follow. Saad et al propose \emph{Contra}, a solution which considers various metrics to identify and evict spam~\cite{saad2020contra}. These consist of defining minimum thresholds for transaction age and fees, beyond which transactions are classified as dust and discarded. There is a tradeoff here: increasing the threshold values can result in false positives (when legitimate transactions are classified as spam) and decreasing it can yield false negatives (when spam transactions are categorized as legitimate). The authors undertake a modelling exercise and predict a maximum accuracy of 60\% in identifying spam. This model has not yet been validated on real-world data. 

The authors make two more noteworthy contributions: they describe how dust attacks may be extended by creating new dust transactions descended from earlier ones. They also recommend increasing block size dynamically during dust attacks to increase the probability that dust transactions are mined and thereby escalating the cost of the attack.

Wang et al. analyse Bitcoin transactions from 2009--2017 to construct \emph{Anti-dust}, a classification model based on the Gaussian distribution~\cite{wang2018anti}. Similar to Contra, the authors set a threshold for the amount of Bitcoins being transacted and any value below the threshold is forwarded to a dust transaction pool, making the mechanism susceptible to false positives and false negatives. If the mempool is not full transactions from the dust pool are moved into it. If the dust transaction pool becomes full, dust transactions are discarded. The authors undertake simulations where they generate 500 legitimate transactions and mix in different amounts of dust transactions. When only genuine transactions are present, transaction validation time is estimated at 200 seconds which increases to 25,000 seconds in the presence of dust, and reduces to 215 seconds when Anti-dust is deployed.

The problems with this filtering approach identified earlier still persist: there is as yet no rigorous definition of spam or dust transactions. False positive rates using these methods are significantly higher than the 1--2\% threshold Baqer et al. discussed. If multiple nodes were to deploy these filters, they may function as inadvertent blacklists. Attackers can modify spam transactions to evade filters. As yet, there is no dynamic filtering solution which adapts to network conditions in real-time to identify and filter spam. Implementing real-time filters and separate pools for spam transactions, will incur computation costs and increased local memory consumption which remain to be evaluated. Moreover, solutions such as adapting a fee-per-output policy or dynamic block sizes require hard forks which can be a contentious process.

In a broader sense, moreover, these solutions are spam-centric. They do not address the general problem of increasing transaction flows and mempool congestion in the network. Our work differs in that our primary focus is local memory consumption. We do not identify or filter spam. Instead, we make the mempool resilient to larger traffic flows. Carbyne is orthogonal to filtering and eviction strategies, and may easily be integrated with these if needed.

Previously, Neonpool~\cite{haq2024neonpool} introduced a novel transaction pool design based on Bloom filter variants, offering a scalable solution for light clients. It enables resource-constrained devices, such as smartphones, browsers, and IoT platforms, to perform full-node functions effectively. In Carbyne, we extend this work by conducting a detailed study on Bitcoin, utilizing counting Bloom filters with datasets three times larger than those used previously. This analysis includes an in-depth examination of mempool architecture and an exploration of the trade-offs among error rates, memory utilization, reprocessing costs, security, and computational overhead. We also make our scheme resilient to DDoS attacks.

\subsection{Threat Model}
\label{sec:threatmodel}
In our scenario, we assume users Alice, Bob, and others operate full nodes connected to the Bitcoin network. Their goal is to contribute to the health and footprint of the network by validating and circulating Bitcoin transactions. We assume the network encounters increasing transaction flows and is also prone to frequent congestion events and DoS attacks. Previous analyses mainly considered DoS attacks on mining pools~\cite{vandervort2014challenges}, Bitcoin exchanges~\cite{wueest2014continued} or the blockchain~\cite{baqer2016stressing}. In our case, our attacker Eve deliberately targets the mempool.

Eve's strategy is to flood the network with transactions and clog the mempool of individual nodes. Her potential goals include delaying transaction processing, increasing fees or negatively impacting user experience. Goals also include publicity stunts or controversy (`griefing' the network). Eve may also use spam to cover for double-spends or sophisticated attacks like the MakerDAO attack detailed earlier.

We assume Eve can generate dust transactions at a rate higher than the throughput of the Bitcoin network and of considerable cumulative size, thereby creating a large backlog of unconfirmed transactions in the mempool. She may accomplish this using sybil accounts and automated scripts. However, Eve does not have any control over Bitcoin users or the ability to directly modify the functioning of their nodes.

Carbyne fundamentally rearchitects the mempool by storing transaction fingerprints instead of complete transactions. We are able to process large transaction flows with a dramatically reduced memory footprint (up to two orders of magnitude less) and very low false positive rates (< 0.5\%).

To clarify, Carbyne does not filter transactions or prevent spam and dust attacks. It does not resolve transaction congestion in blocks or directly affect transaction fees. Carbyne does not protect against attacks leveraging spam to compromise user privacy ~\cite{rahalkar2021summarizing}. However, since Carbyne optimizes mempool storage, it effectively copes with congestion and provides strong resilience against dust attacks. Our experiments, using real-world data, indicate that Carbyne can handle up to three times the maximum traffic loads documented on the Bitcoin network in just 10 MB.
 


Carbyne uses probabilistic data structures which yield false positives or false negatives, resulting in a minute amount of transactions being dropped or reprocessed respectively. However, the effect on transaction propagation in the network is not significant: each node initializes its filters independently using random seeds, which makes it highly unlikely that large numbers of nodes drop the same transactions. With a false positive probability of $p$, the probability that the same transaction is categorized as a false positive by $n$ nodes is $p^n$. As $p$ is less than $10^{-3}$, the probability of more than two nodes categorizing it as a false positive is infinitisemally small. The same argument applies to false negatives.


\begin{figure}
\includesvg[width=\columnwidth]{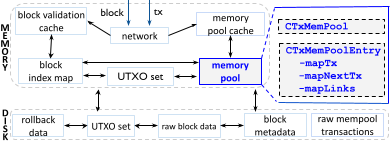}
\caption{Memory and Disk Schematic}\label{fig:schematic}
\end{figure}

\subsection{Bitcoin Mempool and Structure}
\label{sec:mempool}

We briefly describe the Bitcoin mempool and its structure.
%
The mempool is an in-memory repository for unconfirmed transactions and is essential to Bitcoin's transaction propagation mechanism. When a node receives a valid new transaction, it stores it in the mempool while it is being processed by the network. Nodes use this real-time index of transactions to ensure they broadcast incoming transactions over the network only when first received. This minimizes traffic overhead and prevents infinite loops in the network.

The size of the mempool depends on the number of transactions it contains and their individual size, as determined by the transaction content, including the number of inputs and outputs and the length of the scripts. Fig.~\ref{fig:alltime} shows raw transaction data size. Raw transactions of almost 100 MB take up to 300 Mb of memory~\cite{300mbproblem}. Bitcoin's mempool size has varied dramatically over the years, ranging from the order of a few megabytes to over several hundred megabytes, and potentially impacting live transaction fees (Fig.~\ref{fig:txfeebtc}).

Fig.~\ref{fig:schematic} depicts node storage components distributed over hard disk and RAM. Major components in RAM are the \emph{mempool} and the \emph{UTXO set} which stores unspent transaction outputs to validate incoming transactions. 
Other components include 
\emph{memory pool cache} which temporarily stores new transactions pending validation before being accepted to the mempool; 
\emph{block validation cache} which stores signature verification results of unconfirmed transactions to avoid repeating the computationally expensive verification operations when blocks arrive; 
\emph{block index map} which optimizes locating specific block data on disk; and 
\emph{network connections} data about peers, threads, etc.

Major components of disk storage are 
\emph{raw block data} \texttt{(blocks/blk*.dat)} i.e.\ block data that make up the blockchain; 
\emph{block metadata} \texttt{(blocks/index/*)} which optimize the process of locating specific block data; and 
\emph{rollback data} \texttt{(blocks/rev*.dat)} which caters to block reorganization events. These latter databases are redundant as they can be reconstituted from the \emph{raw block data}, but they are maintained on disk to speed up and optimize block validation and other operations~\cite{datastorage}.

Some local memory storage components are also maintained on disk: the entire \emph{UTXO set} \texttt{(chainstate/*)} (over 79 million inputs, 4\,GB in size) stored on disk and partially mirrored in RAM~\cite{dryja2019utreexo}. Similarly, the mempool is maintained in RAM, but since Bitcoin Core version 0.14.0, \emph{raw mempool transactions} are saved to disk at node shutdown, so the Bitcoin mempool state persists across restarts~\cite{btc014}.

Storage within the mempool is governed by two main classes: 1. \texttt{CTxMemPoolEntry} has a bookkeeping role and stores transactions data including size, fee, fee delta, entry height, coinbase status, and information about ancestor and descendant transactions~\cite{txmempoolsourcecode};  2. \texttt{CTxMemPool} is particularly relevant to our study and comprises 3 main structures: 
\texttt{mapTx (boost::multi\_index)}~\cite{boost} sorts the mempool on five criteria: transaction hash, witness-transaction hash, descendant and ancestor fee rate, and time; 
\texttt{mapNextTx (std::map)}~\cite{txmempoolsourcecode} tracks the transaction inputs; and 
\texttt{mapLinks (std::map)}~\cite{txmempoolsourcecode} tracks in-mempool ancestor and descendant transactions.

\texttt{CTxMemPool} and \texttt{CTxMempoolEntry} introduce upto 3x memory overhead in terms of pointers, indexes, and metadata~\cite{300mbproblem}. The minimum recommended storage to set up a Bitcoin full-node is 2\,GB RAM and 350\,GB hard disk space~\cite{fullnode}. The mempool is allocated 300\,MB by default~\cite{mempoollimit}. Users can define a custom mempool acceptance policy. In a low memory environment, it can be reduced (\texttt{-maxmempool}) or disabled entirely (\texttt{-blocksonly}).

\section{Building Blocks: Counting Bloom Filters}
\label{sec:primer}
\begin{figure}[b]
    \centering
    \begin{subfigure}{0.5\textwidth} 
        \includegraphics[width=8.5cm]{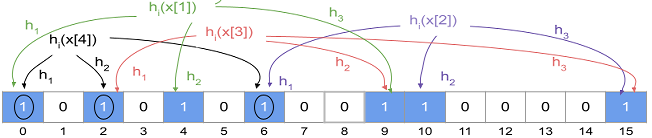}
        \caption{Insertion and queries in a bloom filter} \label{fig:bf}
        \vspace{3mm}
    \end{subfigure}
    \begin{subfigure}{0.5\textwidth} 
      \vspace{3mm}
        \includegraphics[width=8.5cm]{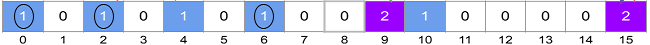}
        \caption{Insertion and queries in a counting bloom filter} \label{fig:cbf1}
         \vspace{3mm}
    \end{subfigure}
    \begin{subfigure}{0.5\textwidth} 
      \vspace{3mm}
        \includegraphics[width=8.5cm]{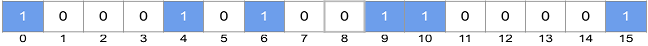}
        \caption{Deletion in a counting bloom filter} \label{fig:cbf}
         \vspace{3mm}
    \end{subfigure}
    \caption{Bloom filters example $(m=16, k=3, n=3)$}
      \label{fig:bfexample}
\end{figure}

\emph{Bloom filters} are memory-efficient probabilistic data structures used to test for set membership~\cite{bloom1970space}. A Bloom filter is essentially a bit vector $v$, of a predefined size of $m$ bits, initially set to zero. To insert an element $x$, belonging to a set $S=\{ x[1], x[2], \ldots, x[n] \}$ of $n$ elements, we hash each $x$ using $k$ independent hash functions with uniformly distributed outputs over the range $\{1,2,\ldots,m\}$, to get $h_1(x)$, $h_2(x)$, $\ldots$, $h_k(x)$, and set the corresponding indices in the bit vector to 1, i.e. $v[h_1(x)] = v[h_2(x)] = \ldots = v[h_k(x)] = 1$. In Fig.~\ref{fig:bf} we insert three elements $x[1], x[2]$, $x[3]$ in the filter, which map to indices $\{2,9,15\}$, $\{6,10,15\}$ \& $\{0,4,9\}$.

To query for set membership, an element $y$ is hashed $k$ times and the corresponding indexes are then checked. If any of the indexes is 0, we can be certain that $y$ is not in the set, a \emph{true negative}. If all relevant indexes are set to 1, $y$ may be in the set. We illustrate this with two examples: If we query the filter regarding $x[1]$ which maps to indices $\{2,9,15\}$, the element was in the set and the bloom filter reported it so, referred to as a \emph{true positive}. If we query the filter regarding $x[4]$, which maps to indexes $\{0, 2, 6\}$, we observe that, even though $x[4]$ was not inserted into the filter, the combination of earlier insertions has set those same bits to 1, resulting in a \emph{false positive}. The probability of false positives, $p$, or \emph{false positive rate (FPR)}~\cite{bose2008false} is:
\begin{eqnarray}
\label{eq:fpr}
p & \approx & \left( 1-e^{-\frac{kn}{m}} \right)^k 
\end{eqnarray}
\emph{FPR} highlights the trade off between space and accuracy. Filter size $m$ can be provisioned as per set size $n$: 
\begin{eqnarray}
\label{eq:size}
m & \approx & n \cdot \frac{-\ln(p)}{(\ln (2))^2} 
\end{eqnarray}
Optimum value of number of hash functions, $k$ is given by:
 \begin{eqnarray}
 \label{eq:hash}
k & \approx &  \frac{m}{n} \cdot \ln(2) 
\end{eqnarray}

\emph{Counting bloom filters} extend bloom filter indices from a single bit to a multi-bit counter (or \emph{bucket}), enabling delete operations. Insert and delete operations increment and decrement the corresponding counters by 1, respectively. For instance, in Fig.~\ref{fig:cbf1}, the counting bloom filter contains elements $x[1], x[2], x[3]$. When $x[3]$ is deleted, index $\{2, 9, 15\}$ are decremented to $\{0,1,1\}$, as shown in Fig.~\ref{fig:cbf}. 

Counting bloom filters also produce \emph{false negatives}. For instance in Fig.~\ref{fig:cbf}, a query for $x[4]$ will result in a false positive. Deleting $x[4]$ will result in indexes $\{0, 2, 6\}$ being decremented. Hence, a subsequent query for $x[1]$ or $x[2]$ will results in a \emph{false negative}. Thus incorrect deletion of the false positive items lead to false negatives. The occurrence of false negatives also increases the false positive rate with time as an item is not deleted, although it should be \cite{guo2009dynamic}. 

\section{Proposed Solution: Carbyne}
\label{sec:proposed}
We now describe the design of Carbyne. Carbyne stores fingerprints of transactions instead of retaining full transactions, whilst preserving essential transaction verification and forwarding functionality. 

Counting bloom filters are commonly used to cache unbounded real-time data streams (for a primer on counting bloom filters, see \S\ref{sec:primer}). However, we face considerable additional challenges as we seek to preserve key functions of the mempool: 
devising a mechanism to expire transactions based on age; 
devising a mechanism for the bulk expiry of transaction inputs; 
tracking and preventing double-spends; and 
ensuring resilience against DoS attacks.

\begin{figure*}[t]
\centering
     \begin{subfigure}{0.24\textwidth}
         \includesvg[width=\textwidth]{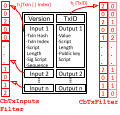}
         \caption{Carbyne Data Structures}
         \label{fig:tx}
     \end{subfigure}
    \hfill
     \begin{subfigure}{0.74\textwidth}
        \includesvg[width=\textwidth]{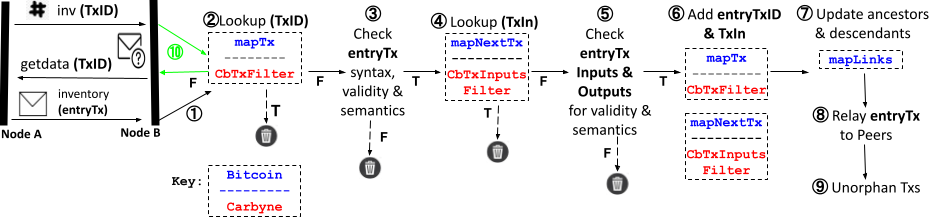}
         \caption{Mempool Entry and Inventory for Bitcoin Core and Carbyne}
        \label{fig:entry}
     \end{subfigure}
     \hfill
      \vspace{-1mm}
      \caption{Entry and Inventory Processes}
       \label{fig:entry1}
\end{figure*}

To address these requirements, we propose a novel design that comprises two counting bloom filter constructions, as shown in Fig.~\ref{fig:tx}: the first filter, \texttt{CbTxFilter} maps valid incoming transactions using the unique transaction hash \texttt{TxID}. Its counterpart in Bitcoin Core is \texttt{mapTx}. The second filter, \texttt{CbTxInputsFilter} checks for duplicate inputs using the tuple \texttt{<TxIn, Index>}, i.e. the \texttt{TxID} of the previous transaction and index of the input within the specified transaction. Its counterpart in Bitcoin Core is \texttt{mapNextTx}. Transactions are removed by deleting the \texttt{TxID} from \texttt{CbTxFilter} and its inputs from \texttt{CbTxInputsFilter}.

\subsection{Entry}
Fig.~\ref{fig:entry} details the mempool entry process and associated structures for Bitcoin Core~\cite{protocolrules} and Carbyne, depicted in blue and red respectively.

\circled{1} A transaction, \texttt{entryTx}, is received over the network via an \texttt{inventory} message. \circled{2} In Bitcoin Core, the \emph{TxID} of the transaction \texttt{entryTx} is used to query \texttt{mapTx} to check if the transaction already exists in the mempool. In Carbyne, \texttt{TxID} is used to query \texttt{CbTxFilter}. An already-received transaction is dropped. \circled{3} If the transaction is new, it undergoes syntax and semantics checks. These checks are identical for both Bitcoin Core and Carbyne and invalid transactions are rejected in either case.

\circled{4} Next inputs, \texttt{TxIn}, of \texttt{entryTx} are scanned for double-spends. In-mempool inputs are verified from \texttt{mapNextTx} in Bitcoin Core and using \texttt{CbTxInputsFilter} in Carbyne. This step was a particular challenge for Carbyne and necessitated the deployment of a separate dedicated filter. If any of the inputs already exist in the mempool, the transaction is dropped. \circled{5} Transaction inputs are also validated using the UTXO set. This check is also identical for Bitcoin Core and Carbyne and transactions with invalid or spent UTXOs are not permitted to enter the mempool. If any transaction input (referred to as \emph{parent} or \emph{ancestor}) is missing, \texttt{entryTx} is added to the orphan pool. It resides in the orphan pool until its ancestor is received.

\circled{6} If \texttt{entryTx} and its inputs are successfully verified, in Bitcoin Core it is added to \texttt{mapTx}, and each of its inputs to \texttt{mapNextTx}. In Carbyne, the \texttt{TxID} is added to \texttt{CbTxFilter} and each of its inputs, \texttt{<TxIn, Index>}, is added to \texttt{CbTx InputsFilter}. \circled{7} In Bitcoin Core, the ancestor-descendant transaction chains are also updated in \texttt{mapLinks}. This information is required by miners to identify and prioritize transactions to be mined in the next block. Carbyne users have the flexibility to store a subset of complete transactions as per their resources, for purposes of mining or to bootstrap new nodes (see detailed discussion in Appendix~\ref{sec:appendix_optionalpool}).

\circled{8} The transaction hash \texttt{TxID} is then broadcast to the node's peers with an \texttt{inv} message and full transactions are forwarded on request. After a short interval to propagate the transaction, Carbyne discards it. The length of the interval that full transactions are stored can vary and may be configured (the trade-off is quantified in Appendix~\ref{sec:appendix_txretention}).

\circled{9} Nodes typically receive \texttt{inv} messages announcing a transaction multiple times. To check if a transaction already exists in the mempool, \texttt{mapTx} in Bitcoin Core and \texttt{CbTxFilter} in Carbyne is queried using \texttt{TxID}. A fresh transaction is requested via the \texttt{getdata} message~\cite{protocolrules}.  

\subsection{Exit}
Transactions exit the mempool for six reasons: 
1.~inclusion in a block, 
2.~the transaction, or one of its unconfirmed ancestors, conflict with a transaction included in a block, 
3.~replacement by a newer version paying more fee, 
4.~eviction due to mempool size limitation, 
5.~a chain reorganization event at the node, and 
6.~expiry due to age. 

\begin{figure}[htbp]
         \centering
         \includesvg[width=0.48\textwidth]{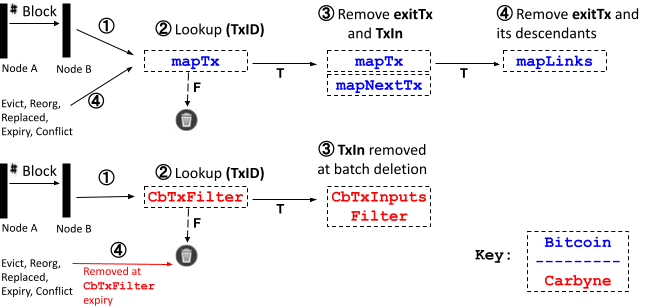}
         \caption{Mempool Exit for Bitcoin Core and Carbyne }
        \label{fig:exit}
\end{figure}

Fig.~\ref{fig:exit} depicts the exit process for Bitcoin Core and Carbyne: \circled{1} When a block arrives over the network, it includes transactions that have been confirmed and should be removed from the mempool. \circled{2} In Bitcoin Core, \texttt{mapTx}, and in Carbyne \texttt{CbTxFilter}, are queried to confirm that the transaction exists in the mempool. \circled{3} In Bitcoin Core, the transaction is removed from \texttt{mapTx}, \texttt{mapNextTx}, and  \texttt{mapLinks}, and in Carbyne, it is removed from \texttt{CbTxFilter} only. 

\circled{4} If the transaction needs to be removed for reasons other than inclusion in a block, Bitcoin Core will remove the transaction as well as its descendants from \texttt{mapTx}, \texttt{mapNextTx}, and \texttt{mapLinks}. In Carbyne, however, these transactions accumulate and are removed by clearing the filter at periodic intervals. For this purpose, we cascade an additional filter to \texttt{CbTxFilter} (as shown in Fig.~\ref{fig:tx}) to work in rotation. This configuration enables us to separate mempool transactions based on age. This process is detailed in \S\ref{sec:results}. 

\texttt{CbTxInputsFilter} is also cleared after a predefined interval, to prevent overflow and degradation in performance. Batch deletion relieves us of the need to store individual mappings of transactions and their inputs, and also saves time over deleting each transaction individually. 

\emph{Replace-by-Fee (RBF)} 
is an opt-in node policy enabling to replace a transaction with a new version by paying a higher fee~\cite{btctxreplacement}. Carbyne does not specifically cater to RBF, but it allows nodes to process and circulate new versions of prior transactions regardless of fee (described in \S\ref{sec:CbTxInputsFilterDynamics}). We discuss potential strategies to enable RBF in Appendix~\textbf{\ref{sec:appendix_rbf}}.

\newtext{
\section{Methodology}
We follow a systematic approach to evaluate Carbyne’s performance against the traditional Bitcoin Core mempool. The methodology consists of six key steps:

\textbf{Data Collection:} We deployed an instrumented version of Bitcoin Core to collect data over a 90-day period, capturing real-time network activity. Our dataset includes mempool activity and statistics along with network activity. The details of the dataset are provided in ~\S\ref{sec:dataset}. This dataset enables a comprehensive reconstruction of the Bitcoin mempool state, serving as the foundation for our evaluation.

\textbf{Design and Implementation:} Carbyne is implemented in C++, the same language as Bitcoin Core, the most widely used Bitcoin client. C++ provides a robust standard library optimized for performance, efficient memory management, and low-overhead data handling. We implement \texttt{CbTxFilter} and \texttt{CbTxInputsFilter}, both counting Bloom filters using the Berkeley libbf library~\cite{libbf}, replacing Bitcoin Core’s \texttt{mapTx} and \texttt{mapNextTx} structures.

\textbf{Experimental Setup:} We conduct our evaluation, running a simulated Bitcoin Core mempool and Carbyne in parallel while processing the recorded dataset. Bitcoin Core’s default mempool serves as the ground truth, while Carbyne represents the experimental condition. Each transaction event, entry, exit, and inventory, is replayed in both environments under identical workloads, allowing a direct performance evaluation.

\textbf{Performance Evaluation:} We evaluate the system based on false positive and negative rates, memory, and computational efficiency. The results are presented through graphs, statistical analysis, and comparative tables to provide a clear and quantitative performance comparison.

\textbf{Discussion and Analysis:} We analyze the system's behavior, linking theoretical expectations to experimental results. This includes an examination of false positives and negatives, a challenge addressed through careful parameter tuning. Additionally, we evaluate the security of our approach, particularly in adversarial conditions, to assess its robustness against potential attacks.

\textbf{Stress Testing for Dos-Resilience:}
To assess Carbyne’s resilience against Denial-of-Service (DoS) attacks, we conduct controlled stress tests designed to evaluate its ability to handle extreme transaction loads while maintaining efficiency and accuracy. We simulate extreme congestion with 600,000 transactions—over three times Bitcoin's peak observed volume. To quantify the impact of congestion on filter accuracy and memory usage, we examine two mitigation strategies in \S\ref{sec:stresstest}.

Our dataset does not specifically include dust or spam transactions, as Carbyne is not a spam filtering mechanism. Instead, it optimizes transaction processing and enhances network capacity without differentiating between legitimate and spam transactions. Since our approach treats all transactions identically, incorporating spam transactions into the dataset would not alter the results. However, Carbyne is complementary to existing spam mitigation techniques and can operate alongside them.}

\section{The \emph{MempoolState} Dataset}
\label{sec:dataset}
To test Carbyne we create an instrumented customized version of Bitcoin Core version 0.11. We used it to set up a full Bitcoin node, running on Intel Core i7-8700 CPU @3.2\,GHz $\times$12, 16\,GB RAM, 2\,TB HDD, running Ubuntu 18.04. We assume that our observation of the network is largely consistent with the rest of the network. Research confirms high similarity in mempools of full nodes~\cite{dae2020examining} and this approach has been widely used in the literature (e.g.~\cite{ali2018zombiecoin}~\cite{baqer2016stressing}). Our dataset is longitudinal, spanning 90 days from 1 January, 2021 to 31 March, 2021 (2160 hours). It comprises: 

\begin{enumerate}
\item 
\emph{Mempool Activity:} We modify Bitcoin Core, specifically \texttt{src/txmempool.cpp}, to capture all mempool entries and exits in JSON format. We log $\sim$29 million unique transactions with $\sim$88 million inputs. This dataset is novel and enables researchers to efficiently reconstruct the Bitcoin mempool state for various applications apart from Carbyne. 

\item 
\emph{Network Inventory:} includes the \texttt{inv} messages received over the network. We log $\sim$89 million \texttt{inv} messages via the \texttt{-network flag} in the Bitcoin configuration file. We store them together with timestamps in CSV format. 

\item 
\emph{Mempool Statistics:} includes the raw transaction size, resulting memory usage, and transaction count in the mempool, obtained using the Bitcoin daemon’s JSON-RPC interface, specifically the \texttt{getmempoolinfo} method. We invoked the \texttt{getmempoolinfo} method at 10-minute intervals and stored the data as JSON. 
\end{enumerate}
 
\section{Experiments and Results}
\label{sec:results} 
In this section we describe our experiments. We undertake a detailed assessment of Carbyne versus the Bitcoin Core mempool for a range of performance metrics, including accuracy, memory consumption, and processing time.

\subsection{Implementation and Methodology} 
We use the \emph{Mempool Activity} and \emph{Network Inventory} data set components to recreate transaction flow at a Bitcoin node over a 90-day period. Each of $entry$, $exit$, and $inv$ is treated as a distinct event that changes the mempool state. We replay these events and process them in parallel,  using Carbyne and a simulated version of the Bitcoin Core mempool. The latter serves as ground truth, enabling us to document precisely how Carbyne processes each event compared to the how Bitcoin Core originally handled it. 

We use C++ to implement Carbyne and simulate the Bitcoin Core mempool. We use the Berkeley libbf library \cite{libbf} to implement counting bloom filters. This library uses the \texttt{H3}${}_\mathtt{exp}$ class of hashing functions designed by Carter and Wegman~\cite{carter1979universal}. The Carbyne implementation consists of probabilistic data structures \texttt{CbTxFilter} and \texttt{CbTxInputsFilter}. We simulate Bitcoin Core mempool's key structures, \texttt{mapTx}, \texttt{mapNextTx} and \texttt{mapLinks}. The \texttt{CbTxFilter} in Carbyne and the \texttt{mapTx} structure in Bitcoin Core are independently queried at every $entry$, $inv$ and $exit$ event in our dataset to check if the relevant transaction exists in the mempool or not. Due to its probabilistic nature, Carbyne's \texttt{CbTxFilter} will sometimes deviate from the ground truth and yield false positives and negatives. We define metrics for these next.


\subsubsection{Performance Metrics}
The responses to mempool queries can be categorized into a confusion matrix: 
\emph{True Positives (TP)}: Bitcoin Core mempool and Carbyne both indicate that the queried transaction is present; 
\emph{True Negatives (TN)}: Bitcoin Core mempool and Carbyne both indicate that the transaction is not in the pool; 
\emph{False Positives (FP)}: Bitcoin Core mempool indicates that the transaction is not in the pool, but Carbyne erroneously records it as present; and 
\emph{False Negatives (FN)}: Bitcoin Core mempool indicates that the transaction is present, but Carbyne erroneously reports it as absent.

At the filter level, the outcomes as per event are as follows: For an $entry$ event, when the mempool is queried on adding a received transaction: 
$\mathrm{TP_{{en}ty}}$: the transaction already exists in the pool and will be discarded; 
$\mathrm{TN_{entry}}$: the transaction is new and will be added to the pool; 
$\mathrm{FP_{entry}}$: the transaction is new and should be added to the pool but will erroneously be discarded; 
$\mathrm{FN_{entry}}$: the transaction already exists in the pool, but will erroneously be added again. 

For an $inv$ event, when the mempool is on a transaction being available at a node: 
$\mathrm{TP_{inv}}$: the transaction already exists in the pool and the full transaction will not be requested; 
$\mathrm{TN_{inv}}$: the transaction is new and the full transaction will be requested to add to the pool; 
$\mathrm{FP_{inv}}$: the transaction is new but will erroneously not be requested to add to the pool; 
$\mathrm{FN_{inv}}$: the transaction already exists in the pool but the full transaction will erroneously be requested to add to the pool.

At $exit$, when the mempool is queried to remove a transaction: 
$\mathrm{TP_{exit}}$: the transaction exists in the pool and will be removed; 
$\mathrm{TN_{exit}}$: the transaction does not exist in the pool and cannot be removed; 
$\mathrm{FP_{exit}}$: the transaction does not exist in the pool but is erroneously `removed' by decrementing the filter counters; 
$\mathrm{FN_{exit}}$: the transaction exists in the pool but is erroneously not removed.

Each filter-level outcome has different consequences for Carbyne performance. False positives at all three events lead to transactions not being processed and hence a reduction in the overall accuracy of the system. Therefore, an overall false positive rate is a vital metric to consider. Specifically, any false positives at inventory and entry result in transactions being discarded and the rate of such discarding needs to be kept down. Looking at false negatives, any such outcome at inventory (or equivalently entry) will cause unnecessary reprocessing of transactions. False negatives at exit will not inflict any immediate cost to the system, but will eventually cause more false positives and hence their effect can be captured by the overall false positive rate. Based on these insights, we define performance metrics for Carbyne:

\begin{itemize}
\item 
\emph{False Positive Rate (FPR)} is a measure of accuracy, defined as the ratio of the false positives to the total number of queries (\emph{entry}, \emph{inv} and \emph{exit}).
\[\mathrm{
FPR=\frac{FP_{entry} + FP_{inv} + FP_{exit}}{Queries_{entry}+ Queries_{inv}+ Queries_{exit}} 
}\]

\item 
\emph{Discarded Transactions} is a measure of the proportion of new transactions at $entry$ and $inventory$ that were erroneously discarded due to false positives.
\[\mathrm{
DiscardedTxs=\frac{FP_{inv}+FP_{entry}}{Queries_{inv} + Queries_{entry}}
}\]

\item 
\emph{Reprocessed Transactions} is a measure of transactions that were processed twice due to false negatives at $inventory$.
\[\mathrm{
ReprocessedTxs=\frac{FN_{inv}}{Queries_{inv}}. 
}\]
\end{itemize}

We note here that there is precedent for erroneous transaction handling in the Bitcoin ecosystem, especially for resource-constrained clients. For instance, false positives in SPV clients occasionally result in forwarding of erroneous transactions. However, circulating these transactions does not equate to actual double-spends, since nodes in the network, including Carbyne nodes, will still screen transactions in all incoming blocks to ensure there is no double-spending.



\subsubsection{Dimensioning \texttt{CbTxFilter}}
The highest transaction volumes observed to date have edged close to the 200k transactions mark, as depicted in Fig.~\ref{fig:alltime}. We therefore choose a maximum transaction load of 200k as a starting point to dimension \texttt{CbTxFilter} and a false positive rate of $10^{-3}$ (1 in 1000). This is an order of magnitude less than the 1\% failure rate Baqer et al. suggest may be `disruptive' and `a denial of service in itself'~\cite{baqer2016stressing}. Using Eq.~\ref{eq:size} and Eq.~\ref{eq:hash}, we derive a starting bloom filter size of 600\,KB consisting of 2.4M buckets and 8 hash functions.

\begin{table*}[t]
\small
\setlength{\tabcolsep}{4pt}
\begin{tabular}{rrrrrrrrrrrrrrr}
\toprule
\multicolumn{1}{c}{\textbf{\texttt{CbTx}}} & \multicolumn{1}{c}{\textbf{Bu-}}    & \multicolumn{1}{c}{\textbf{Hash}}       & \multicolumn{3}{c}{\textbf{False Positive Rate}} & \multicolumn{2}{c}{\textbf{Discarded Transactions}}& \multicolumn{2}{c}{\textbf{Reprocessed Transactions}} \\
\multicolumn{1}{c}{\textbf{\texttt{Filter}}} &                 \multicolumn{1}{c}{\textbf{ckets}}    & \multicolumn{1}{c}{\textbf{Func.}}  & \multicolumn{1}{c}{\textbf{Theore-}} & \multicolumn{1}{c}{\textbf{No Expiry}} & \multicolumn{1}{c}{\textbf{Expiry}} & \multicolumn{1}{c}{\textbf{No Expiry}}   & \multicolumn{1}{c}{\textbf{Expiry}} & \multicolumn{1}{c}{\textbf{No Expiry}}  & \multicolumn{1}{c}{\textbf{Expiry}} \\
\multicolumn{1}{c}{\textbf{size}} & \multicolumn{1}{c}{\textbf{\emph{(m)}}} & \multicolumn{1}{c}{\textbf{\emph{(k)}}} & \multicolumn{1}{c}{\textbf{tical}} & \multicolumn{1}{c}{} & \multicolumn{1}{c}{}                    & \multicolumn{1}{c}{\textbf{Num/(\%)}} & \multicolumn{1}{c}{\textbf{Num/(\%)}} &     \multicolumn{1}{c}{\textbf{Num/(\%)}}  & \multicolumn{1}{c}{\textbf{Num/(\%)}} \\
\midrule
\rowcolor[HTML]{E7E6E6}  %
\textbf{600 KB} & 2.4M & 8 & 1.2E-03 & 1.5E-02 & 3.7E-03 &  1,646,878(1.402) & 497,795(0.339) & 82,028(0.1964) & 30,034(0.0339) \\
\textbf{800 KB} & 3.2M & 11 & 4.6E-04 & 9.3E-03 & 1.4E-03 & 1,050,915(0.893) & 195,073(0.133) & 58,063(0.0655) & 10,186(0.0115) \\
\rowcolor[HTML]{E7E6E6}
\textbf{1\,MB}   & 4M   & 14 & 6.7E-05 & 8.1E-03 & 8.1E-04 & 911,695(0.774) & 109,338(0.074) & 52,650(0.0594) & 5,822(0.0066) \\
\textbf{2\,MB}   & 8M   & 28 & 5.0E-09 & 3.8E-03 & 1.3E-04 & 427,242(0.362) &  17,441(0.012) & 30,712(0.0346) &  803(0.0028) \\
\rowcolor[HTML]{E7E6E6} 
\textbf{3\,MB}   & 12M  & 42 & < 1E-10 & 2.3E-03 & 4.9E-05 & 263,263(0.223) &  7,096(0.005)  & 22,608(0.0255) &  298(0.0003) \\
\textbf{4\,MB}   & 16M  & 55 & < 1E-10 & 1.4E-03 & 2.8E-05 & 163,923(0.139) &  4,273(0.003)  & 15,600(0.0176) &   77(0.0001) \\
\midrule  
\end{tabular}
\vspace{-2mm}
  \caption{Performance Metrics for \texttt{CbTxFilter} of various sizes dimensioned for $n=200,000$ transactions}
  \label{tab:cbtxfilter}
\end{table*}

\begin{figure*}[t]
    \centering
    \begin{subfigure}{0.33\textwidth} 
        \includegraphics[width=\textwidth]{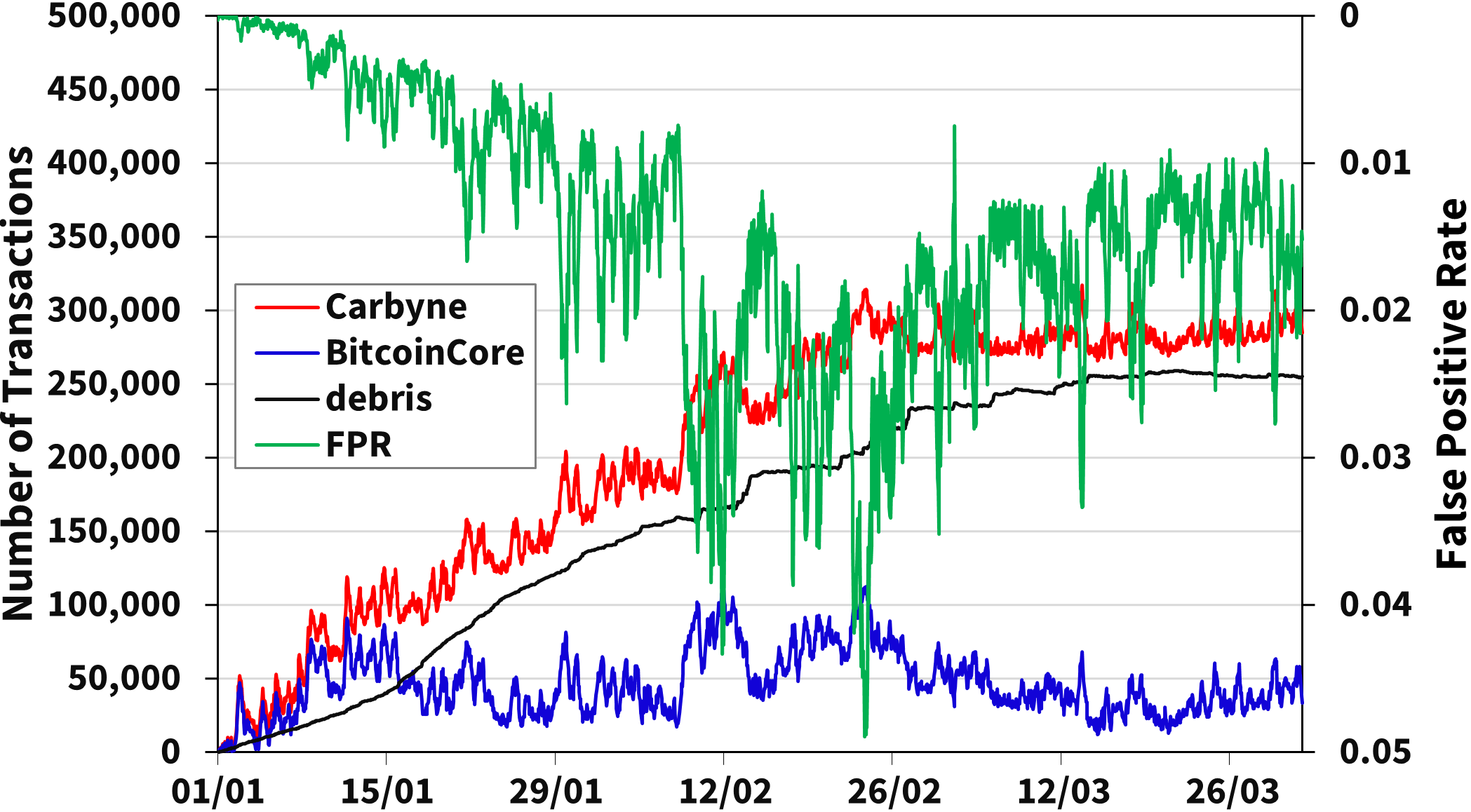}
        \caption{\texttt{CbTxFilter}= 600\,kB}
        \label{fig:600nex}
   \end{subfigure}
   \hfill
    \begin{subfigure}{0.325\textwidth} 
        \includegraphics[width=\textwidth]{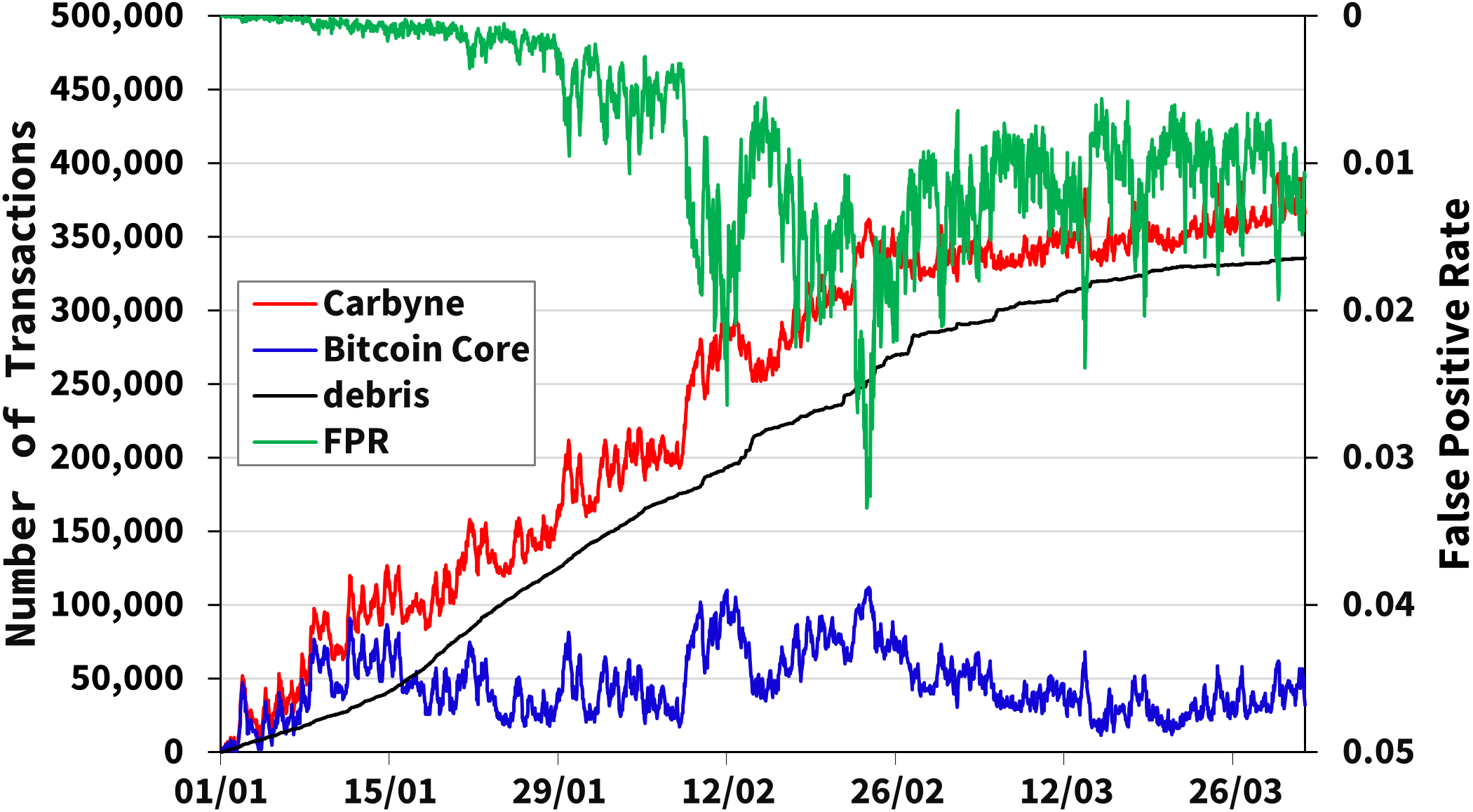}       
        \caption{\texttt{CbTxFilter}= 1\,MB}
        \label{fig:1nex}
    \end{subfigure}
    \hfill
     \begin{subfigure}{0.33\textwidth} 
        \includegraphics[width=\textwidth]{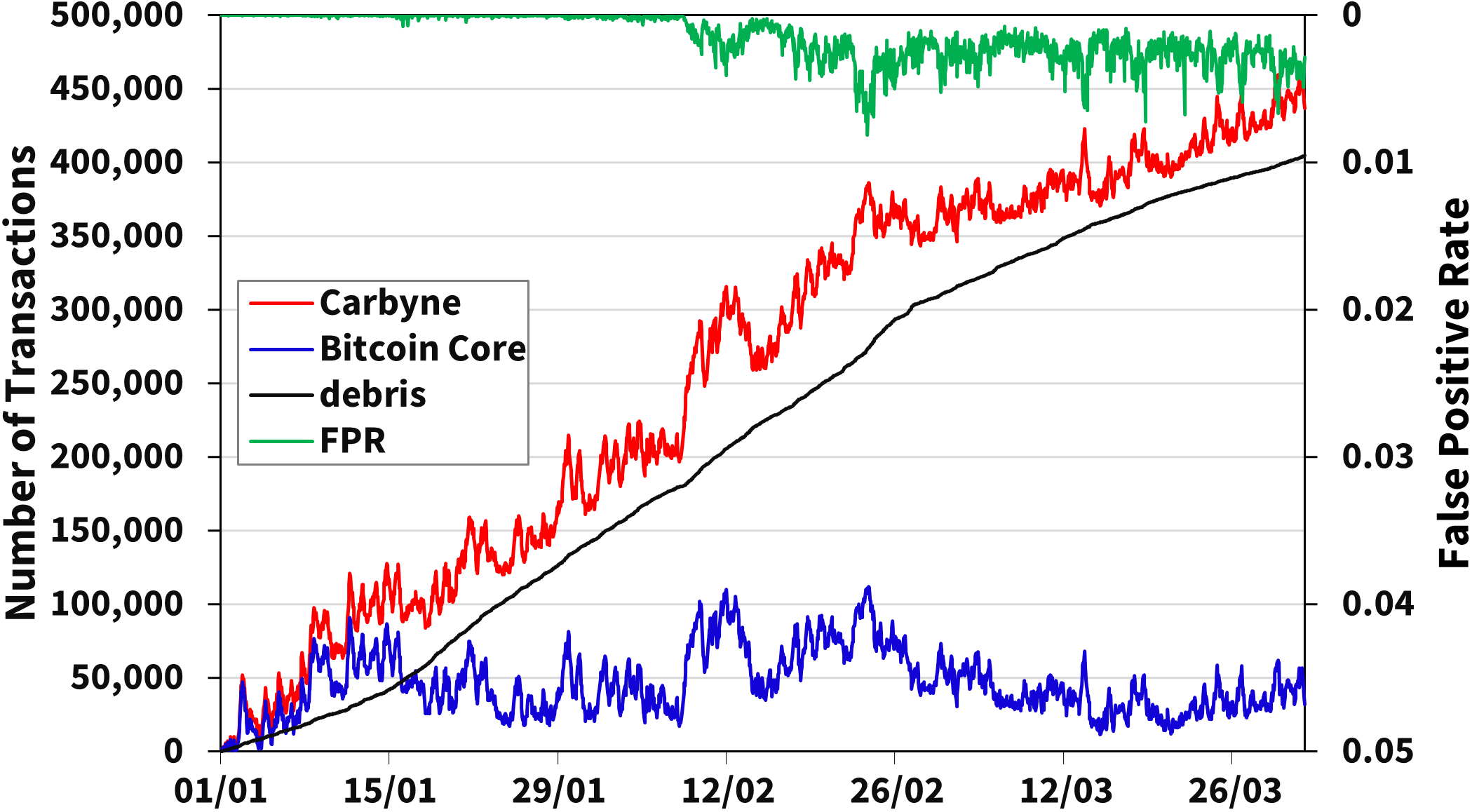}
          
        \caption{\texttt{CbTxFilter}= 4\,MB}
        \label{fig:4nex}
    \end{subfigure}
    \hfill
    \begin{subfigure}{0.33\textwidth} 
    \vspace{2mm}
        \includegraphics[width=\textwidth]{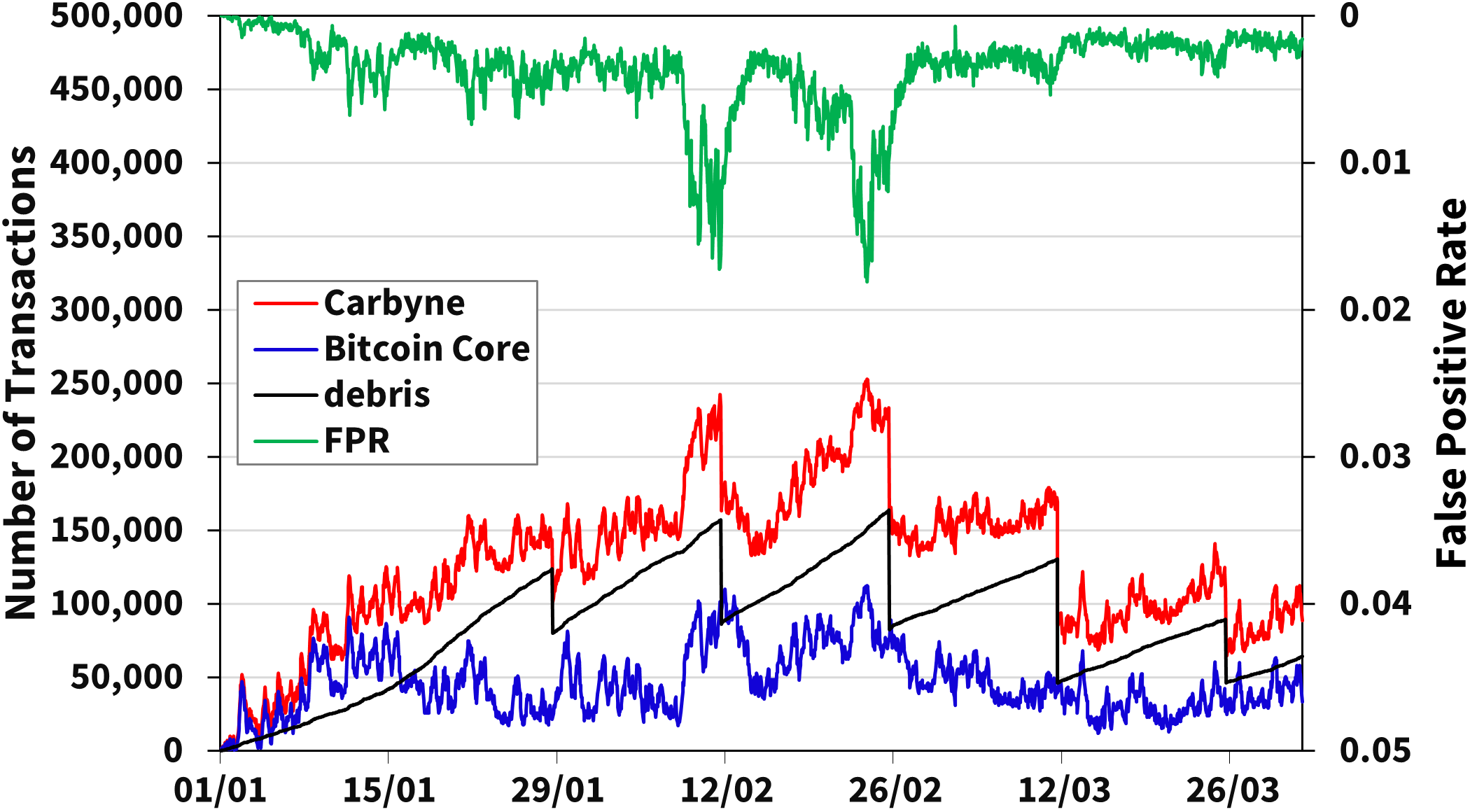}
        \caption{\texttt{CbTxFilter}= 600\,kB$\times$2}
        \label{fig:600ex}
   \end{subfigure}
   \hfill
    \begin{subfigure}{0.325\textwidth} 
    \vspace{2mm}
        \includegraphics[width=\textwidth]{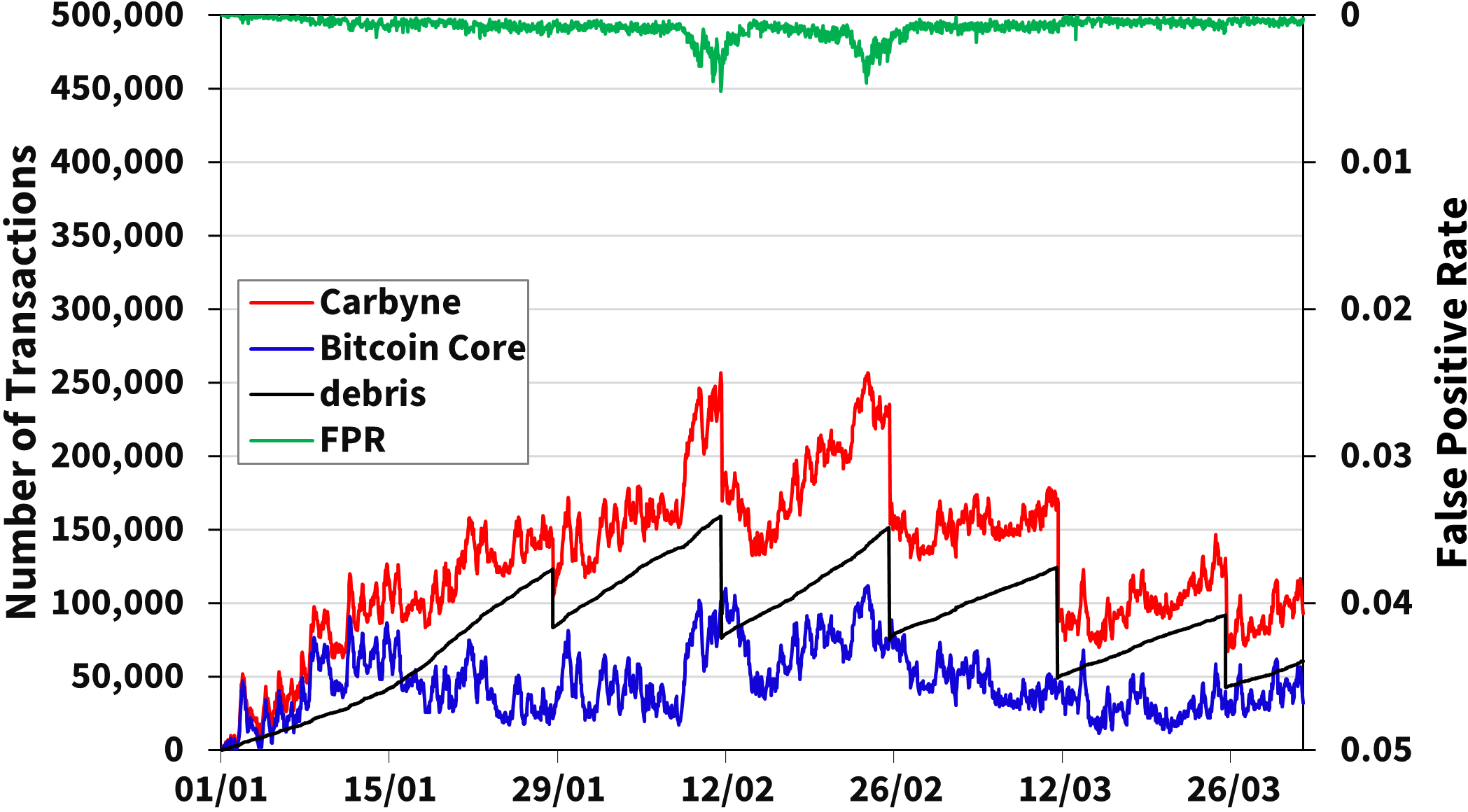}
        \caption{\texttt{CbTxFilter}= 1\,MB$\times$2 }
        \label{fig:1ex}
    \end{subfigure}
    \hfill
     \begin{subfigure}{0.33\textwidth} 
     \vspace{2mm}
        \includegraphics[width=\textwidth]{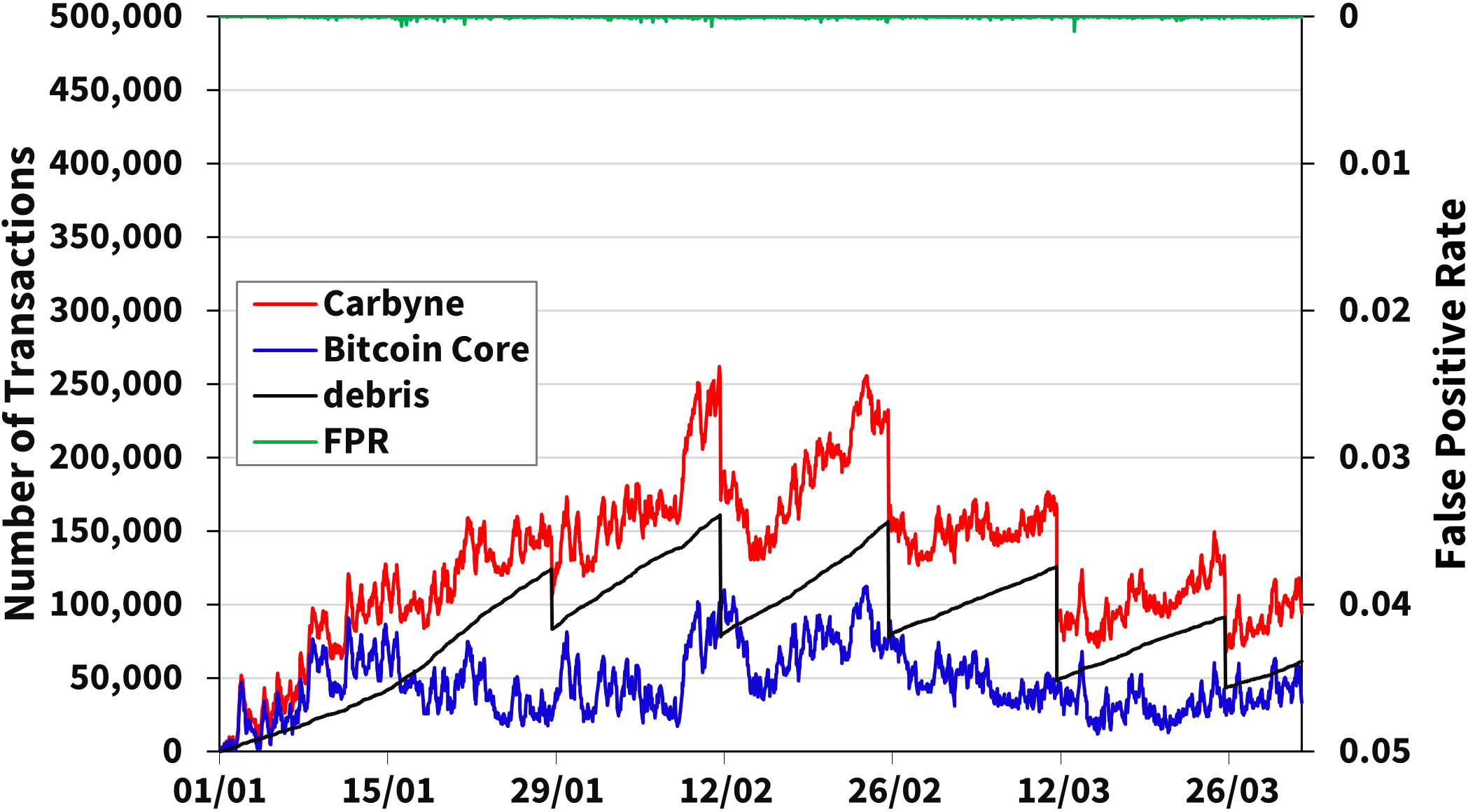}
        \caption{\texttt{CbTxFilter}= 4\,MB$\times$2 }
        \label{fig:4ex}
    \end{subfigure}
    \vspace{-2mm}
    \caption{Number of Transactions in Bitcoin vs Carbyne, Debris Transactions and False Positive Rate}
     \label{fig:nexex}
    \hfill
\end{figure*}

We also test other bloom filter sizes, listed in Table~\ref{tab:cbtxfilter}. For comparative purposes, we focus on two other filter candidates for our experiments, sized at 2\,MB (medium) and 4\,MB (large), with 8M and 16M buckets each, and provisioned with 14 and 28 hash functions respectively. Fan et al. recommend a bucket width of 4 at which probability of overflow in counting bloom filters is infinitesimally small~\cite{fan2000summary}. However, in our experiments, counters in the filters did not exceed 2. We therefore allocate a bucket width of 2 bits.

\subsection{Empirical Results and Discussion}
We replay transaction events in the \emph{MempoolState} dataset for the three month period through all filters. Table~\ref{tab:cbtxfilter} shows the average false positive rate is highest for the 600\,kB filter at 1.46\E{-2}, reducing marginally to 8.07\E{-3} for a 1\,MB filter, and further to 1.43\E{-3} for the 4\,MB filter. For the 600\,kB, 1\,MB, and 4\,MB filters we also observe 1.402\%, 0.774\%, and 0.139\% of transactions being erroneously \emph{discarded} due to false positives and 0.1964\%, 0.0594\%, and 0.0176\% of transactions being \emph{reprocessed} due to false negatives. These trends are as expected with false positives and negatives decreasing with increasing filter size.

Fig.~\ref{fig:nexex} depicts in real-time the number of transactions in these filters along with the false positive rate for the entire three month period. We also plot the corresponding number of transactions in the Bitcoin Core mempool, the ground truth in our evaluation. In all three cases, we observe the number of transactions closely tracks the pattern in the Bitcoin Core mempool, with an increasing offset. 

This offset is due to two main sources: first, transactions that should have been removed from the mempool because of age or were replaced or in conflict with other transactions. The Bitcoin Core mempool `expires' old transactions after a two-week (default) period and may evict certain transactions if they are in-conflict with any in-mempool transactions. Counting bloom filters have no inherent mechanism to track age or conflicts. Second, false negatives result in some transactions being erroneously added to the mempool at \emph{entry} events and some, due to be removed, to persist at \emph{exit} events. We term these erroneous artifacts \emph{debris}, which accumulates over time and corresponds to growing deterioration in performance.

We also observe in Fig.~\ref{fig:600nex}--\ref{fig:4nex}, a marked increase in the false positive rate in all three filters, after the number of transactions reach 200,000, i.e. the size the filters were originally dimensioned for. For the 600\,kB and 1\,MB filters, transaction numbers for Carbyne and Bitcoin Core do not diverge indefinitely but stabilize around the 300,000 mark. Our investigation reveals this is due to the high false positive rate, which has the interesting effect of evicting large numbers of transactions from the filter at \emph{exit} events (by decrementing the counters). This, in turn, significantly reduces the transactions count and the false positive rate.

This is similar to expiry in temporal counting bloom filters, where counters are periodically decremented by a decay factor to prevent saturation of the filters for continuous streams of item insertions~\cite{zhao2013design}. This effect does not occur in the 4\,MB filter, because the false positive rate is too low to evict large numbers of transactions, and the offset between Carbyne and Bitcoin Core continues to grow.

Empirical false positive rates are also significantly higher than our theoretical target of $10^{-3}$ for all filters, almost by an order of magnitude (e.g.\ 6.7\E{-5} vs.\ 8.1\E{-4} for the 1\,MB filter). We theorize as to the causes: first, multiple works have reported false positive rates in real deployments are higher than theoretically computed~\cite{mullin1983second}~\cite{gremillion1982designing}. Researchers contend that this is because theoretical calculations assume that ``each hash transformation is perfect''~\cite{mullin1983second} and that transactions ``are independent and uniformly distributed over all records'' whereas real activity tends to be ``clumped''~\cite{gremillion1982designing}. In this context, Bose et al. prove that Eq.~\ref{eq:fpr} actually gives us a lower bound on the false positive rate~\cite{bose2008false}.

Second, repeated insertion and deletion from an unbounded set over time, leads to increased false positives and false negatives. Moreover, debris rapidly causes transactions in \texttt{CbTxFilter} to exceed the 200,000 mark for which it was provisioned. This results in a complex dynamic and interdependent effects that compound over time. For instance, Guo et al. note in their study that false negatives at \emph{exit} will cause items that should be removed to persist in the filter, in turn resulting in yet more false positives~\cite{guo2010false}. We can confirm that we observe this specific effect in our own experiments.

\subsubsection{Transaction Expiry}
There is no inherent mechanism to resolve the issue of debris in counting bloom filters. We propose a solution to periodically `clean up' the filters. This is done by employing two identical counting bloom filters, a primary and a secondary, working together in rotation, which switch status after a predefined interval. This configuration enables us to clearly separate transactions on the basis of age.

\begin{table*}[ht]
\small
\setlength{\tabcolsep}{2.5pt}
\begin{tabular}{crrrrrrrrr}
\midrule
\multicolumn{1}{c}{\textbf{Expiry}} & \multicolumn{3}{c}{\textbf{False Positive Rate}} & \multicolumn{3}{c}{\textbf{Discarded Transactions}} & \multicolumn{3}{c}{\textbf{Reprocessed Transactions}} \\
\multicolumn{1}{c}{\textbf{Time}} & \multicolumn{1}{c}{\textbf{600 kB}} & \multicolumn{1}{c}{\textbf{1 MB}} & \multicolumn{1}{c}{\textbf{4 MB}} & \multicolumn{1}{c}{\textbf{600 kB}} & \multicolumn{1}{c}{\textbf{1 MB}} & \multicolumn{1}{c}{\textbf{4 MB}} & \multicolumn{1}{c}{\textbf{600 kB}} & \multicolumn{1}{c}{\textbf{1 MB}} & \multicolumn{1}{c}{\textbf{4 MB}} \\
\multicolumn{1}{c}{\textbf{\emph(days)}} & \multicolumn{3}{c}{\textbf{}} & \multicolumn{1}{c}{\textbf{Num/(\%)}} & \multicolumn{1}{c}{\textbf{Num/(\%)}} & \multicolumn{1}{c}{\textbf{Num/(\%)}}  & \multicolumn{1}{c}{\textbf{Num/(\%)}} & \multicolumn{1}{c}{\textbf{Num/(\%)}} & \multicolumn{1}{c}{\textbf{Num/(\%)}} \\
\midrule
\rowcolor[HTML]{E7E6E6} 
\textbf{14--28}                       & 3.7E-03 & 8.1E-04 & 2.8E-05 & 497,795(0.339) & 109,338(0.074) & 4,273(0.003) &  30,034(0.034) &   5,822(0.007) &     77(\,<\,0.001)  \\
\textbf{7--14}                        & 2.1E-03 & 4.6E-04 & 2.2E-05 & 287,670(0.196) &  62,271(0.042) & 3,519(0.002) &  34,549(0.039) &  18,169(0.021) &  16,562(0.019) \\
\rowcolor[HTML]{E7E6E6} 
\textbf{1--2}                         & 6.2E-04 & 1.5E-04 & 1.6E-05 &  85,589(0.058) &  20,814(0.014) & 2,654(0.002) & 148,128(0.167) & 126,017(0.142) & 146,265(0.165) \\
\textbf{$\mathbf{\frac{1}{2}}$--1}             & 4.4E-04 & 1.2E-04 & 1.5E-05 &  60,852(0.042) &  16,258(0.011) & 2,537(0.002) & 231,543(0.261) & 191,210(0.216) & 231,428(0.261) \\
\rowcolor[HTML]{E7E6E6} 
\textbf{$\mathbf{\frac{1}{4}}$--$\mathbf{\frac{1}{2}}$} & 3.1E-04 & 8.3E-05 & 1.4E-05 &  43,085(0.030) &  11,790(0.008) & 2,400(0.002) & 294,164(0.332) & 262,567(0.296) & 294,530(0.280) \\ 
\midrule
\end{tabular}
\vspace{-2mm}
\caption{False positive rates for different Expiry times and \texttt{CbTxFilter} sizes, dimensioned for n=200,000 transactions}
\label{tab:expirytime}
\end{table*}

All queries are directed to the primary filter first. If a transaction is not present there, then the secondary filter is queried. Transactions are removed from the filter that first reports them to be present. However, insertions only occur into the primary filter. After every predefined interval, the secondary filter is reset, i.e. all counters are decremented to zero, and status of the two filters is switched again. This cycle repeats after every interval, and effectively simulates Bitcoin's transaction expiry mechanism, where transactions are automatically removed from the mempool after a default period of 14 days.

We implement this mechanism with an expiry period of 14 days. Results are in Table~\ref{tab:expirytime} and Figs.~\ref{fig:600ex}--\ref{fig:4ex}. The filters first switch roles on 15 January, followed by the first expiry event on 29 January, and then again every 14 days, corresponding to sharp drops in transaction numbers in the filters. The Carbyne mempool now more closely tracks the contours of the Bitcoin Core transactions pattern. Average false positive rates for our three filters of interest are reduced by at least one order of magnitude, to the point that performance of filters with expiry is comparable to that of larger filters. For instance, the 1\,MB filter, with expiry demonstrates an average false positive rate of 8.1\E{-4}, which is less than 1.4\E{-3} resulting from the 4\,MB filter without expiry. The number of discarded and reprocessed transactions are also dramatically reduced.

The expiry period here is not strictly 14 days but in the range of 14--28 days, since a filter is cleared every 28 days. The false positive rate can be further improved by reducing the expiry period. A case could be made for this given that, since 2016, the median transactions confirmation time for Bitcoin has not exceeded 30 mins~\cite{median-confirmation-time} and the average confirmation time of a transaction is around 104 mins ~\cite{avg-confirmation-time}.

This means that even 6--12 hours is a plausible value for expiry. As shown in Table ~\ref{tab:expirytime}, when the expiry period is reduced from 14--28 days to $\frac{1}{4}$--$\frac{1}{2}$ day, there is an order of magnitude decrease in the false positive rate for the 600\,kB and 1\,MB filters which reduced from 3.7\E{-3} to 3.1\E{-4} and 8.1\E{-4} to 8.3\E{-5}, respectively. For the 4\,MB filter the decrease is only marginal. Similarly, the discarded transactions percentage for the 600\,kB  and 1\,MB filters reduced from $0.339\%$ to $0.030\%$ and from $0.074\%$ to $0.008\%$, respectively. However shorter expiry intervals result in increased reprocessing costs: if an old transaction is received again after the filter has been cleared, Carbyne will consider it a new transaction and process it again. For a 600\,kB filter, reprocessed transactions increase from $0.034\%$ to $0.332\%$, for a 1\,MB filter from $0.007\%$ to $0.296\%$, and for a 4\,MB filter from less than 1\E{-5}\% to $0.280\%$.

\subsubsection{\texttt{CbTxInputsFilter} Dynamics}
\label{sec:CbTxInputsFilterDynamics}
\texttt{CbTxInputsFilter} scans inputs of incoming transactions to prevent double spends. 
The implications are: 
\begin{itemize}
\item 
$\mathrm{TP_{inputs}}$: a transaction bearing that input was recently added to \texttt{CbTxInputsFilter} and the new transaction should be discarded; 
\item 
$\mathrm{TN_{inputs}}$: a transaction bearing that input does not exist in \texttt{CbTxInputsFilter} and the new transaction should be added; 
\item 
$\mathrm{FP_{inputs}}$: a transaction bearing that input does not exist in \texttt{CbTxInputsFilter} but the new transaction was erroneously discarded; 
\item 
$\mathrm{FN_{inputs}}$: a transaction bearing that input already exists in \texttt{CbTxInputsFilter}, but the new transaction was erroneously added.
\end{itemize}

\begin{table}
\small
\centering
\begin{tabular}{rrrr}
\toprule
\textbf{\texttt{CbTx}} & \textbf{Buckets} &\multicolumn{2}{c}{\textbf{False Positives}}  \\
 \textbf{\texttt{Inputs}} & \textbf{\emph{m}} & \multicolumn{1}{c}{\textbf{1 hour}} & \multicolumn{1}{c}{\textbf{3 hour}}  \\
\textbf{\texttt{Filter}} & & \textbf{Num/(FPR)} & \textbf{Num/(FPR)} \\
\midrule
\rowcolor[HTML]{E7E6E6} 
\textbf{600KB} & \textbf{2.4M} & 69,644(7.9E-04) & 446,476(5.1E-03) \\
\textbf{800 KB} & \textbf{3.2M} & 26,665(3.0E-04) & 152,865(1.7E-03) \\
\rowcolor[HTML]{E7E6E6} 
\textbf{1 MB} & \textbf{4M} & 15,258(1.7E-04) & 91,480(1.0E-03) \\
\textbf{2 MB} & \textbf{8M} & 2,249(2.6E-05) & 20,554(2.3E-04) \\
\rowcolor[HTML]{E7E6E6} 
\textbf{3 MB} & \textbf{12M} & 781(8.9E-06) & 10,312(1.2E-04) \\
\textbf{4 MB} & \textbf{16M} & 411(4.7E-06) & 6,128(7.0E-05) \\
\midrule
\end{tabular}
\vspace{-2mm}
\caption{\texttt{CbTxInputsFilter}, n=200,000}
\label{tab:cbtxinputfilter}
\end{table}

In our dataset incoming transactions average 40,000 inputs hourly and peak at 191,947 inputs. We therefore dimension \texttt{CbTxInputsFilter} for 200,000 transactions, similar to \texttt{CbTxFilter}. We reset \texttt{CbTxInputsFilter} hourly. This relieves the requirement of tracking individual inputs to transactions. With hourly reset and \texttt{CbTxInputsFilter} of 600\,kB, 1\,MB and 4\,MB the false positive rate is 7.9\E{-4}, 1.7\E{-4} and 4.7\E{-6} respectively as shown in Table \ref{tab:cbtxinputfilter}.

As we do not individually delete transactions from \texttt{CbTx InputsFilter}, no false negatives occur due to incorrect deletion of false positive items. False negatives occur due to expiry of transaction inputs. Thus an attacker can resend a transaction with the same input after the expiry interval. With regards to transactions that have conflicting inputs there is no hard and fast policy as to which transaction nodes accept. Typically, nodes accept and forward the first seen transaction and the first seen can differ for nodes. Moreover, Core clients with disabled mempools forward all valid incoming transactions. It is the job of miners to not add conflicting transactions to a block and the network nodes to screen incoming blocks. Since Carbyne maintains complete UTXO information, nodes will reject blocks that include double-spend transactions.

For a 600\,kB \texttt{CbTxFilter} (14-28 day expiry) discarded transactions stand at 0.387\%, i.e. 99.613\% of transactions are accurately processed. With a 600\,kB \texttt{CbTxInputsFilter}, an additional 69,644 transactions are discarded due to false positives. For a 1\,MB \texttt{CbTxInputsFilter}, 15,258 transactions are discarded. If we factor this in for a 1\,MB \texttt{CbTxFilter} (14-28 day expiry), the discarded transactions stand at 0.0849\%, i.e. 99.915\% of transactions are accurately processed.  For a 4\,MB \texttt{CbTxInputsFilter} (14-28 day expiry), 411 transactions are discarded due to false positives. Thus for a 4\,MB \texttt{CbTxFilter} discarded transactions stand at 0.003\%, that is  99.997\% transactions are accurately processed.

\begin{figure}
        \includegraphics[width=8.5cm]{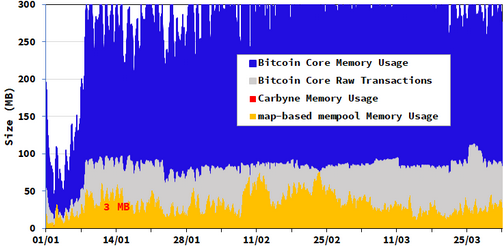}
       \vspace{-1mm}
        \caption{Raw transaction size and memory footprint}
        \label{fig:memusage}
\end{figure}

\subsubsection{Memory and Computation Footprint}
We compare here the memory footprint and computational overhead of Carbyne and Bitcoin Core. 
Fig.~\ref{fig:memusage} depicts the size of the raw transactions versus the actual size of the mempool over the three month period of our experiments. This information is derived from the Mempool Statistics component of our dataset, described in \S\ref{sec:dataset}. As depicted, raw transaction data of 100 MB typically has a footprint of 300 MB or more in the mempool due to data structure overheads to optimize functionality (as noted in \S\ref{sec:background}). The graph clips at 300 MB, i.e. the maximum size allocation of our node and the default allocation for Bitcoin Core. 

As baseline, we consider a straightforward mempool optimization scheme, which uses deterministic data structures but only keeps transaction hash in memory and with fewer indices. Rudimentary calculations indicate that the storage size for this scenario would be significantly more than Carbyne. For complete transaction validation, we would need to store the following indices: transaction ID (32 bytes), transaction input hash (32 bytes) and index (4 bytes) - amounting to $(32+36n)$ bytes for each raw transaction where $n$ is the number of transaction inputs. Using standard data structures, such as \texttt{maps}) would incur a memory overhead almost three times the size of the raw data. Overall, we get savings of up to one order of magnitude. 

This approach reduces memory consumption from an optimization perspective, but the disadvantages are still significant. It is less resilient to congestion events and spam attacks. This node does not store full transactions, it is equally limited as Carbyne from an inventory perspective, i.e., it cannot bootstrap mempools of other nodes. The only benefit is that it does not suffer from false positives.

Carbyne results thus far are immensely promising: to process an equivalent volume of transactions with 99.9151\% accuracy, Carbyne requires only 3 MB of memory, corresponding to \texttt{CbTxFilter} sizes of 1\,MB$\times$2 (with expiry) and a \texttt{CbTxInputsFilter} size of 1\,MB. \newtext{Carbyne represents a two-order-of-magnitude reduction compared to Bitcoin Core, with memory usage dropping from 300 MB to just 3 MB. Notably, this value remains constant regardless of transaction volume. Given this significant efficiency gain, we characterize Carbyne as "ultra-light." This improvement is quantitatively illustrated in Fig.~\ref{fig:memusage}.}

We have identified various parameters users can tweak to navigate the tradeoff between accuracy, memory footprint, and computation overhead as per their own requirements. Filters can be dimensioned based on expected transaction flows and expiry can be tweaked, enabling smaller filters to achieve accuracy comparable to much larger ones.

We next calculate the computational overheads. Bitcoin Core components \texttt{mapTx}, \texttt{mapNextTx} and \texttt{mapLinks} undertake query, insertion and deletion operations in $O(\log n)$ time, where $n$ is the number of stored transactions, 
Counting bloom filters in Carbyne operate in constant time, $O(k)$, for query, insertion and deletion operations, where $k$ is the number of hash functions.

\begin{figure}
\includegraphics[width=8cm]{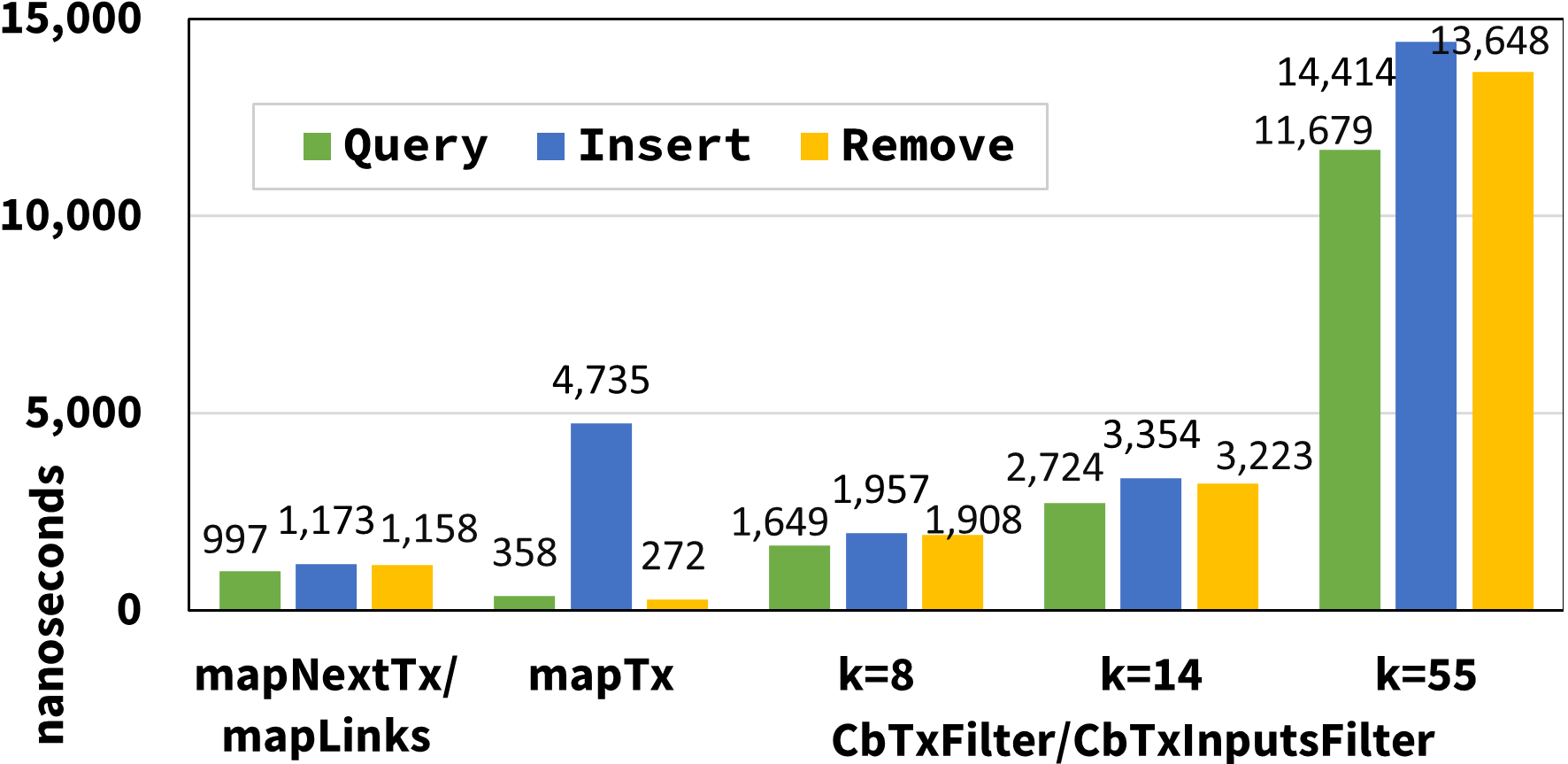}
\vspace{-1mm}
\caption{Computation Time}
\label{fig:timing}
\end{figure}

We compare their timings, depicted in Fig.~\ref{fig:timing}, using an Intel Core i7 system with 8700 CPU @3.2GHZ $\times$12 and 32 GB RAM, running Ubuntu 18.04 with GCC 5.4.0. We replicate Bitcoin Core structures \texttt{mapTx} (Boost multi-index), \texttt{mapNextTx} and \texttt{mapLinks} (C++ STL maps). We use \texttt{libbf} to instantiate counting bloom filters and perform query, insert and delete operations, averaging over 1 million iterations. 

\begin{itemize}
\item 
\textbf{\texttt{CbTxFilter vs. mapTx:}} Query, insert and delete operations in \texttt{mapTx} take 358\,ns, 4,735\,ns, and 272\,ns, respectively, whereas \texttt{CbTxFilter} (for $k=14$) requires 2,724\,ns, 3,354\,ns, and 3,223\,ns, respectively. For a complete transaction lifecycle comprising query, insert and removal, \texttt{mapTx} requires 5,365\,ns compared to 9,301\,ns for \texttt{CbTxFilter}.

\item 
\textbf{\texttt{CbTxInputFilter vs. mapNextTx:}} Query, insert and delete operations for \texttt{mapLinks} and \texttt{mapNextTx} require 997\,ns, 1,173\,ns, and 1,158\,ns, respectively, whereas \texttt{CbTxInputsFilter} (for $k=14$) takes 2,724\,ns and 3,354\,ns for query and insert operations. There is no delete operation for this filter as it is periodically reset (as described in \S\ref{sec:proposed}). Over a single transaction input life-cycle, \texttt{mapNextTx} requires 3,328\,ns compared to 6,078\,ns for \texttt{CbTxInputsFilter}.

\item 
\textbf{\texttt{mapLinks:}} Maintaining unconfirmed transaction chains in \texttt{mapLinks} is memory and computation intensive. For each entry, ancestor-descendent information is kept in the mempool to prioritize transactions for blocks. But with every entry or exit in the mempool, all ancestors and descendants need to be recursively updated. A single series of operations requires 3,328\,ns. To cap these costs, also be a potential DDoS vector, Bitcoin Core 0.12 introduced a default policy limiting unconfirmed chains to 25 transactions and total size of 101\,kB. Carbyne does not store these mappings as explained in \S\ref{sec:proposed} and does not incur these costs.
\end{itemize}

\newtext{
\noindent
For a complete transaction lifecycle—including query, insertion, and removal—\texttt{mapTx} in Bitcoin Core requires 5,365 ns, whereas \texttt{CbTxFilter} incurs a slightly higher cost of 9,301 ns. Considering the transaction lifecycle, \texttt{mapNextTx} in Bitcoin Core requires 3,328 ns, while \texttt{CbTxInputsFilter} in Carbyne demands 6,078 ns. Managing unconfirmed transaction chains within \texttt{mapLinks} over a single sequence of operations requires 3,328 ns in Bitcoin Core. Typical Bitcoin transactions have around 2–3 inputs. Assuming the addition of a 2 input transaction necessitates two sets of query, insertion and removal operations, the total cost in \texttt{mapLinks} amounts to 6,656 ns. In contrast, Carbyne eliminates this overhead entirely. Summing these computational costs, the total processing time per transaction lifecycle amounts to 15,349 ns for Bitcoin Core and 15,379 ns for Carbyne. These figures indicate that Carbyne and Bitcoin Core exhibit comparable computational loads.  

From a practical standpoint, Carbyne is demonstrably capable of supporting real-time transactions. Given that each transaction requires 15,379 ns, the system can theoretically sustain a transaction throughput of approximately 65,000 transactions per second (tps). This significantly exceeds Bitcoin’s current throughput of 3--7\,tps, highlighting Carbyne’s scalability potential. If Bitcoin's throughput increases, Carbyne can seamlessly scale to accommodate loads to the order of thousands of transactions per second.}

\newtext{As we noted, there are no financial incentives in the Bitcoin protocol for full nodes (as compared to miners), and the resource costs for this exercise are constantly increasing. On our part, we assume that users operate nodes primarily to contribute to the Bitcoin network out of a sense of community or altruism -- similar to how people operate nodes for the Tor network.

In this sense, we believe it is useful to provide such users, via our solution, with control over the local memory resources they allocate for this task to make it cost-efficient. This is conceptually similar to Bitcoin Core’s inbuilt ‘pruned node’ option which aims to reduce hard disk requirements. i.e., pruned nodes do not store the entire blockchain on disk but still contribute to the footprint and health of the network by validating and forwarding transactions.

Carbyne, as a lightweight solution, lowers the entry cost for participation in the Bitcoin network. Similar to pruned nodes, Carbyne should be viewed as a solution not to supplant mining nodes but to reduce costs for certain users while enhancing network health by validating transactions and improving congestion resilience. In summary, Carbyne minimizes memory costs while preserving computational efficiency, making it a practical, cost-effective solution for Bitcoin node operators.}

\subsection{Security Analysis in an Adversarial Setting}
Here, we consider if an adversary can exploit Carbyne's design or properties in unintended ways. We specifically discuss two situations: can an adversary craft transactions to evade or trigger filters at individual Carbyne nodes? If so, can the adversary trigger censorship of select transactions across large numbers of Carbyne nodes in the network?

To recap the fundamental point in \S\ref{sec:threatmodel} each node initializes its filters independently using random 128-bit seeds that are kept secret. This makes it highly unlikely that large numbers of nodes drop the same transactions. The seeds we are using to initialize the filters are 128 bits long and kept secret by the nodes (as explicitly recommended in the literature [1]). This means that individual nodes are effectively using independent hash functions to construct their filters. 

To trigger a false positive in a certain Carbyne node, an adversary would therefore have to access two pieces of information: 1) the seeds used to initialize that node’s most recent filter, and 2) the current state of that filter (i.e. which transactions it has already logged). We assume the first item would be difficult for an attacker to procure. A node can even change its random seeds every time a new \texttt{CbTxFilter} is generated (every two weeks in our basic scenario). 

Seeding \texttt{CbTxFilter} with secret random seeds effectively reduces the adversary’s capability from mounting a chosen transaction attack to a random transaction attack: although the adversary controls the transactions they broadcast, without knowledge of the seeds, the output of the hash function and hence the marked indices in the filters will be effectively unpredictable for the adversary. Hence, the adversary cannot do much better than broadcasting random transactions. He will not be able to effectively craft specific transactions that lead to false positives with higher probability. 

Moreover, since individual nodes use random seeds, even if an adversary somehow crafts a transaction that induces a false positive in a targeted node, the probability of inducing a false positive in another node using the same transaction will be very very low (as described in \S\ref{sec:threatmodel}). Adversarial attempts targeted at individual nodes will, therefore, not scale to Carbyne nodes network-wide. We do not believe there is any low-cost means for an attacker to censor specific transactions across all Carbyne nodes on the network.

\section{\emph{Carbyne for Other Cryptocurrencies}} 
\label{sec:othercrypto}
The Carbyne approach, in principle, can be extended to other cryptocurrencies. However the design of such a scheme has to be tweaked considerably according to the specific protocols. Adapting Carbyne to Bitcoin forks such as Bitcoin Cash, Bitcoin Gold, Litecoin and Dogecoin is straightforward, as they are all based on the UTXO model. For these currencies only certain parameters such as transaction expiry time, bloom filter size, etc. need to be tweaked according to the traffic on the network and the block time. 

\subsection{\texttt{CarbEth} (Carbyne for Ethereum)}
An accounts-based cryptocurrency like Ethereum will necessitate considerable changes in addition to the changes identified above. We propose an equivalent scheme for Ethereum, \texttt{CarbEth} (Carbyne for Ethereum). However, the empirical analysis regarding the various parameters users can tweak to navigate the tradeoff between accuracy, memory footprint, and computation overhead will have to be determined based on a dataset specific to the Ethereum network. This investigation warrants a separate and dedicated examination. We will address this in our future research.

\subsection{Ethereum txpool vs Bitcoin mempool}
The transaction pool in Ethereum, namely the \texttt{Txpool} comprises three primary map structures: \texttt{\textbf{pending}} holds executable transaction data; \texttt{\textbf{queued}} manages non-executable transaction data and \texttt{\textbf{beats}} keeps a record of the most recent heartbeat from each known account.

As outlined in \S\ref{sec:mempool}, the Bitcoin mempool is composed of two primary structures: the \texttt{CTxMemPoolEntry}, which plays bookkeeping role, and the \texttt{CTxMemPool}, which encompasses three key data structures: \texttt{mapTx} indexes transactions on five criteria: transaction hash, witness-transaction hash, descendant and ancestor fee rates, and timestamp; \texttt{mapNextTx} tracks transaction inputs; \texttt{mapLinks} monitors in-mempool ancestor and descendant transactions.

\subsection{Primary components of CarbEth \& Carbyne}
CarbEth is composed of two components: The first component, a counting bloom filter named \texttt{pendingFilter}, utilizes the transaction hash (\texttt{<txHash>}) to map and track valid incoming transactions. In Carbyne, the first component, a counting bloom filter named \texttt{CbTxFilter} maps valid incoming transactions using the unique transaction hash \texttt{<TxID>}. 

In CarbEth, the second component a bloom filter with key-value storage, named  \texttt{accountFilter} ensures that double-spend transactions are identified. It does this by storing key-value pair \texttt{<address, nonce>}, which represent the sender's address and the nonce value of the most recent transaction added to \texttt{pendingFilter}. In Carbyne, the second component, a counting bloom filter, \texttt{CbTxInputsFilter} checks for duplicate inputs using the tuple \texttt{<TxIn, Index>}.

\begin{figure*}
         \centering
         \includesvg[width=0.75\textwidth]{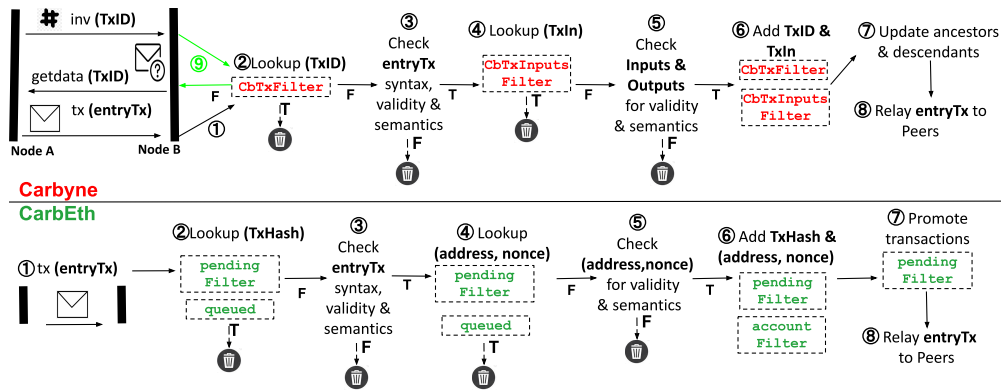}
         \caption{Carbyne vs CarbEth}
        \label{fig:carbeth}
\end{figure*}

\textbf{Entry} 
\circled{1} In Ethereum the process begins with the arrival of a complete transaction, referred to as \texttt{entryTx}, over the network through a \texttt{Transaction} message. In Bitcoin, a transaction announcement which only includes the transaction hash is made via the inv message and a full transaction is relayed only on request. This means that the Bitcoin node has to hold on to transactions, in anticipation of a complete transaction request (quantitative analysis offered in Appendix~\ref{sec:appendix_txretention}). Ethereum can forego this. \circled{2} In CarbEth the transaction's \texttt{txHash} is used to query the \texttt{pendingFilter} and \texttt{queued} data structures to determine if the transaction already exists in the txpool.  In Bitcoin Core, the transaction hash of \texttt{entryTx}, the \texttt{TxID} is used to query \texttt{CbTxFilter} to check the same. \circled{3} If the transaction is new, it undergoes syntax and semantics checks in both CarbEth and Carbyne. 

\circled{4} Next, in CarbEth, the <address, nonce> of \texttt{entryTx} is scanned in the \texttt{accountFilter} to detect any potential double spends.  If a transaction exists in \texttt{pendingFilter} that has the same \texttt{<account, address>} as the incoming transaction, it is flagged as a potential double-spends and dropped. In Carbyne, the inputs, \texttt{TxIn}, of \texttt{entryTx} are scanned for double-spends using \texttt{CbTxInputsFilter} in Carbyne. 

\circled{5} In CarbEth transactions are also validated against the State Trie using the key-value pair \texttt{<address, nonce>}. In Carbyne, transaction inputs are also validated using the UTXO set. If any transaction input (referred to as \emph{parent} or \emph{ancestor}) is missing, \texttt{entryTx} is added to the orphan pool.  

\circled{6}  Upon successful verification, in CarbEth, the \texttt{txHash} is added to \texttt{pendingFilter}, and the key-value pair \texttt{<address, nonce>} is added to \texttt{accountFilter} if the nonce is in order, or to \texttt{queued} if the nonce is out of order. In Carbyne If the \texttt{entryTx} and its inputs are successfully verified, the \texttt{TxID} is added to \texttt{CbTxFilter} and each of its inputs, \texttt{<TxIn, Index>}, is added to \texttt{CbTxInputsFilter}. 
\circled{7} In CarbEth the node then attempts to promote any eligible transactions from \texttt{queued} to \texttt{pending}. In Bitcoin Core, the ancestor-descendant transaction chains are updated and transactions are unorphaned. 
\circled{8} In CarbEth the complete transaction is then relayed to a small, random fraction of connected peers via a \texttt{Transaction} message. The transaction hash \texttt{TxID} is then broadcast to the node's peers with an \texttt{inv} message and full transactions are forwarded on request. This entry process is summarized in Fig. \ref{fig:carbeth}.

\textbf{Exit} transactions are removed from the txpool for several reasons: inclusion in a block, replacement by higher fee versions, reaching size limits, removal during chain reorganization, transaction age expiry. Additionally in Ethereum transactions may be removed for running out of gas during execution, and demotion to ``queued'' (non-executable) status. To remove transactions in \texttt{CarbEth}, the \texttt{txHash} is deleted from \texttt{pendingFilter}, and the corresponding \texttt{<address, nonce>} entry in \texttt{accountFilter} is updated. In Carbyne, the transaction is removed from the \texttt{CbTxFilter} and the \texttt{CbTxInputsFilter} is cleared periodically.

Nonetheless, it is imperative to empirically validate the suggested approach using Ethereum-specific data. We require accurate measurements and quantification of outcomes, particularly concerning the rates of false positives and negatives and to understand various parameters through evidence-based analysis.


\begin{figure*}[t]
    \centering
    \begin{subfigure}{0.33\textwidth} 
        \includegraphics[width=\textwidth]{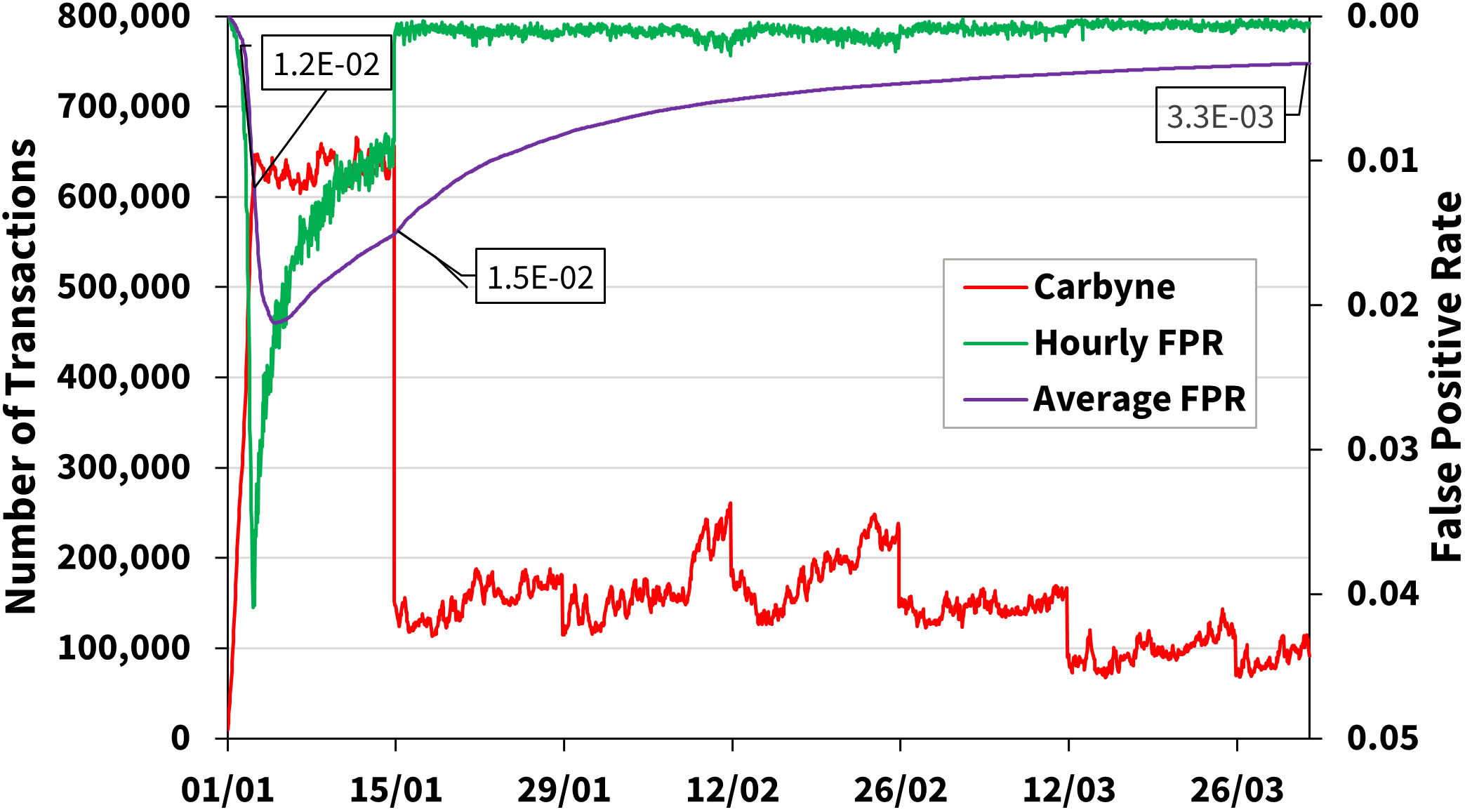}
        \caption{\texttt{CbTxFilter}= 1.8\,MB$\times$2}
        \label{fig:lar1}
   \end{subfigure}
   \hfill
    \begin{subfigure}{0.325\textwidth} 
        \includegraphics[width=\textwidth]{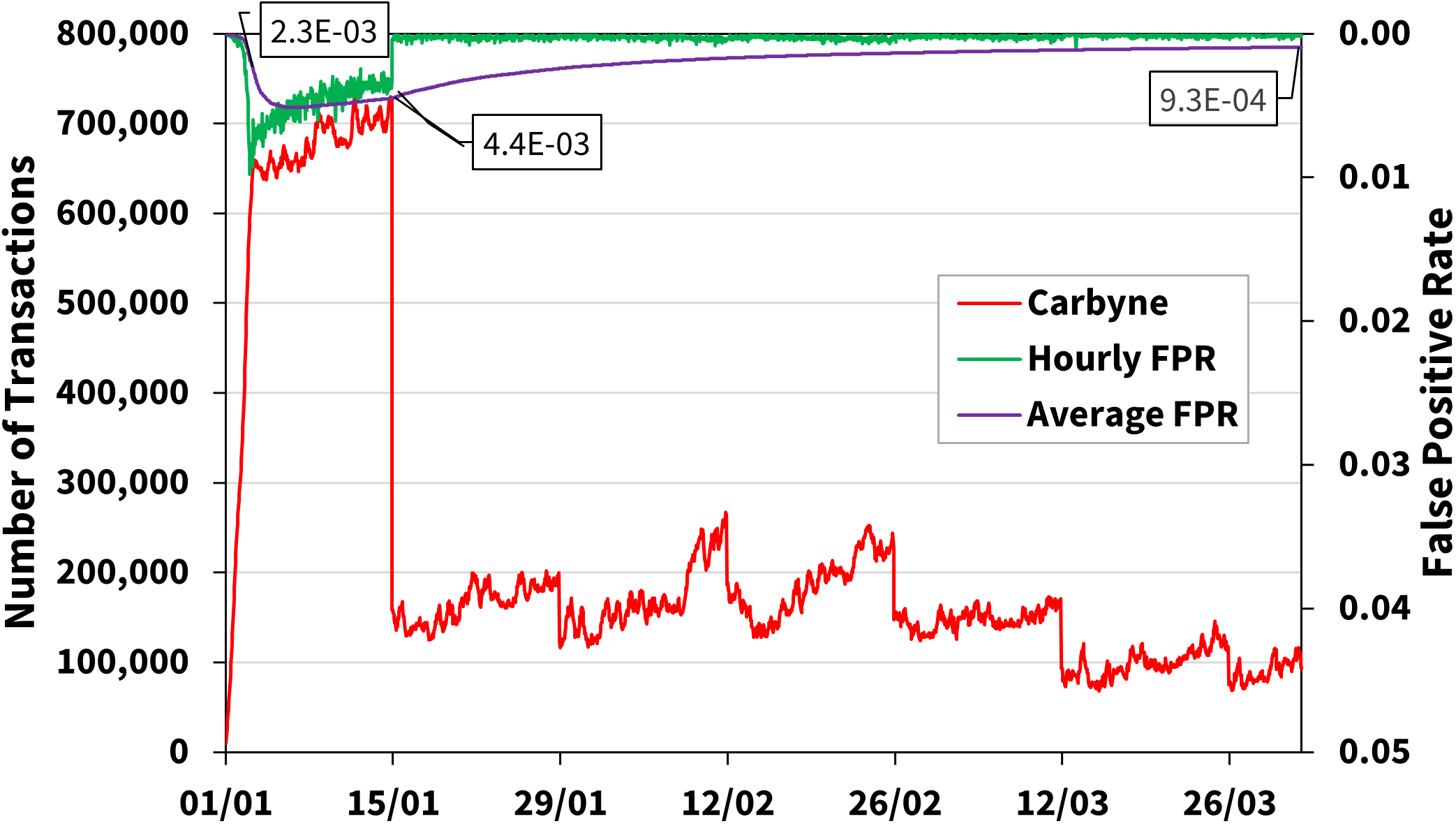}
        \caption{\texttt{CbTxFilter}= 3\,MB$\times$2}
        \label{fig:lar2}
    \end{subfigure}
    \hfill
     \begin{subfigure}{0.33\textwidth} 
        \includegraphics[width=\textwidth]{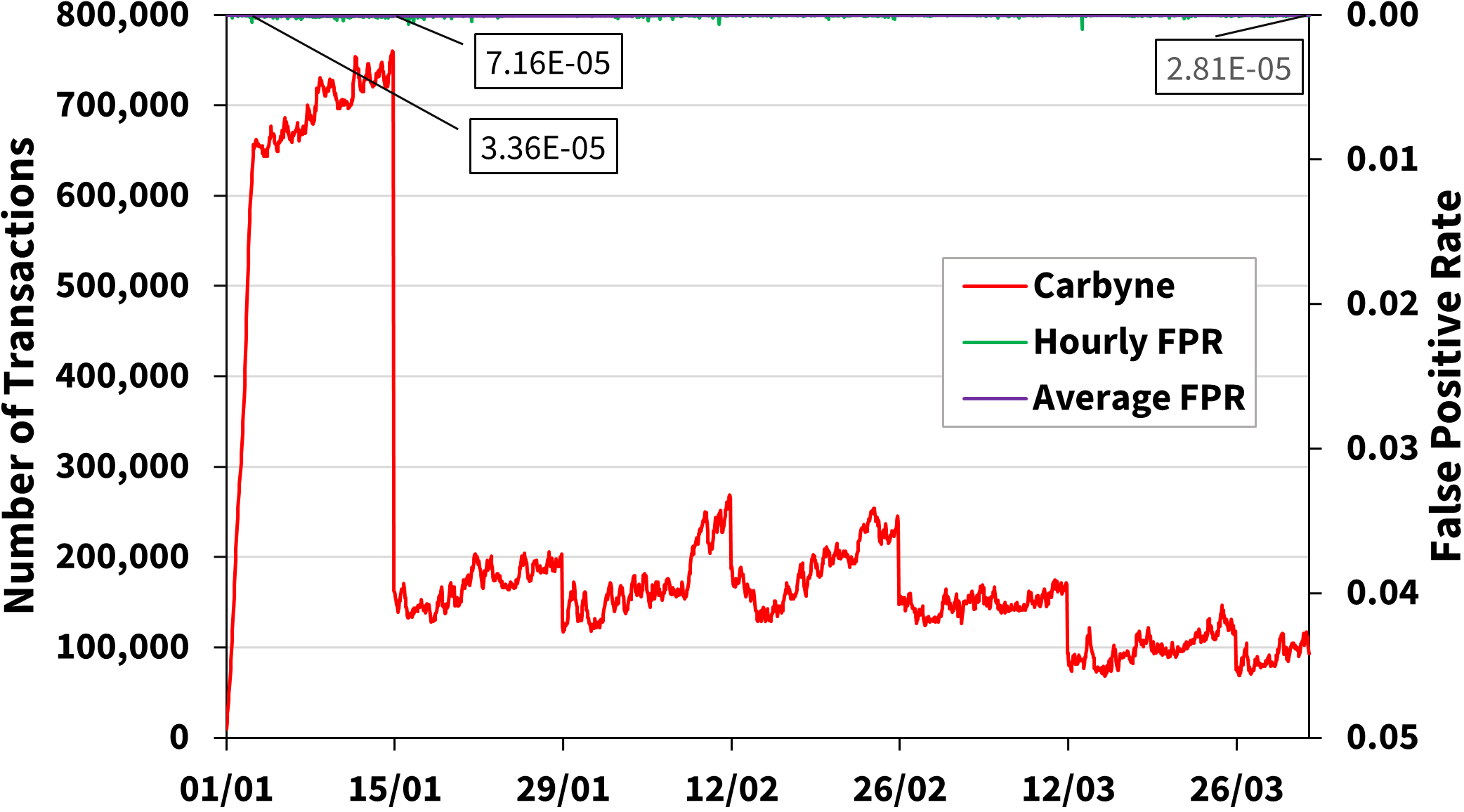}
        \caption{\texttt{CbTxFilter}= 12\,MB$\times$2}
        \label{fig:lar3}
    \end{subfigure}
    \hfill
        \centering
    \begin{subfigure}{0.33\textwidth} 
        \includegraphics[width=\textwidth]{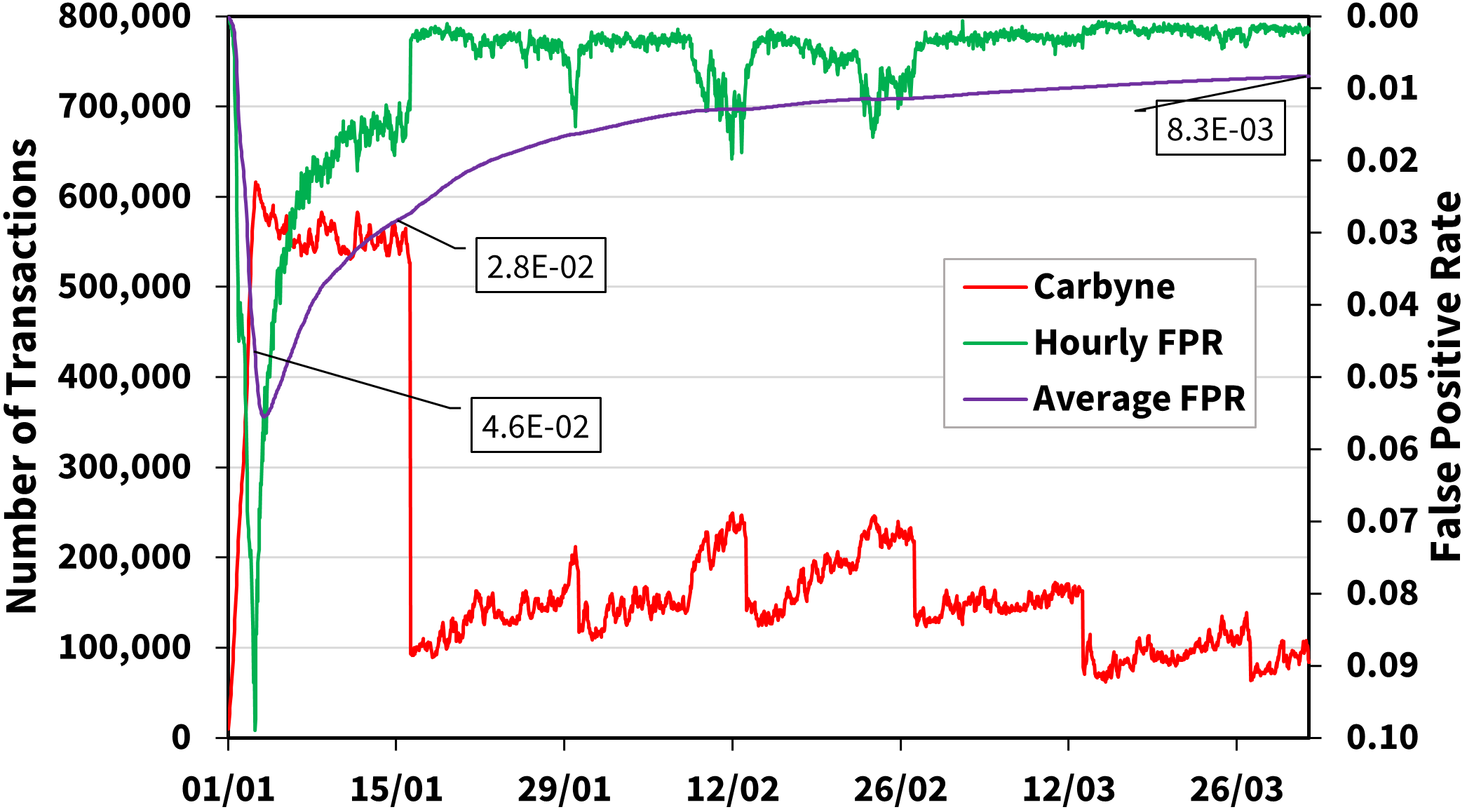}
        \caption{\texttt{CbTxFilter}= 600\,kB$\times$4}
        \label{fig:dyn1}
   \end{subfigure}
   \hfill
    \begin{subfigure}{0.325\textwidth} 
        \includegraphics[width=\textwidth]{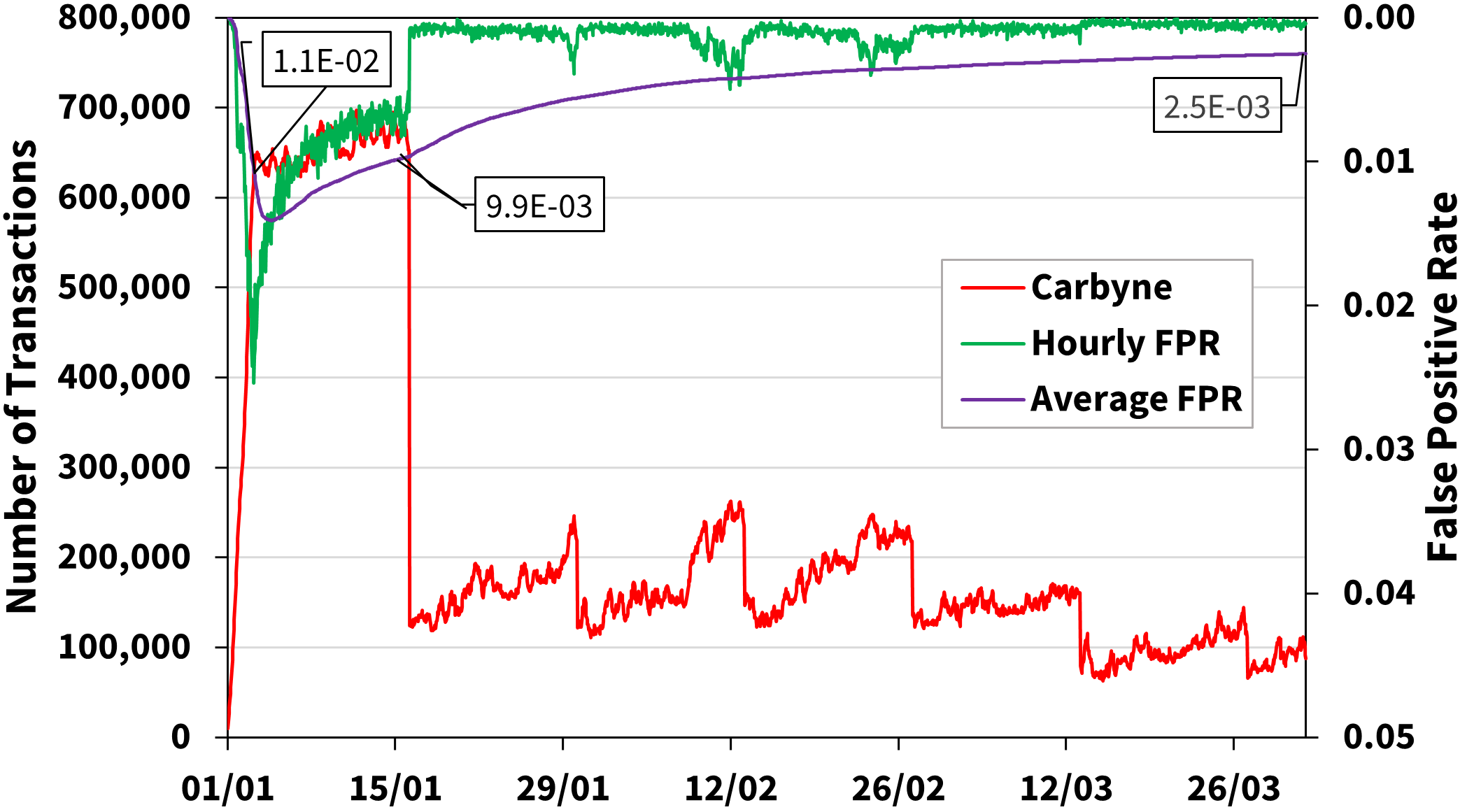}
        \caption{\texttt{CbTxFilter}= 1\,MB$\times$4}
        \label{fig:dyn2}
    \end{subfigure}
    \hfill
     \begin{subfigure}{0.33\textwidth} 
        \includegraphics[width=\textwidth]{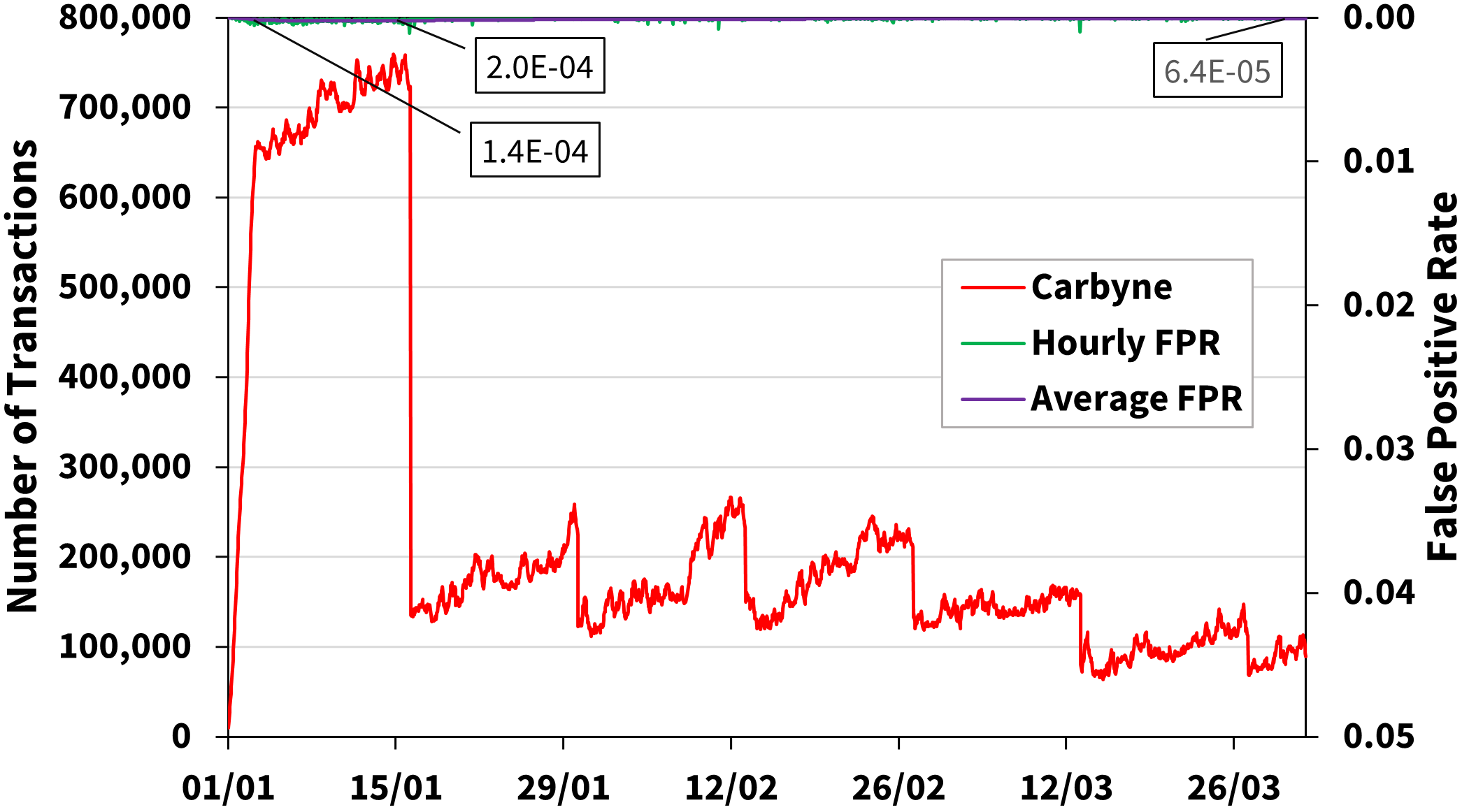}
        \caption{\texttt{CbTxFilter}= 4\,MB$\times$4}
        \label{fig:dyn3}
    \end{subfigure}
    \vspace{-2mm}
    \caption{Dimensioning for larger transaction volume and dynamic adaptation for congestion and DDoS}
     \label{fig:stress}
    \hfill
\end{figure*}

\section{Stress Testing Carbyne}
\label{sec:stresstest}
We simulate a stress test to evaluate Carbyne's performance under high transaction flows, congestion or spam attacks. We consider a transaction load of over 600,000 transactions, more than 3x the peak amount considered in \S\ref{sec:results}. This load is generated by artificially halting exit of transactions for 55 hours to let entry transactions accumulate. Once 600,000 transactions accumulate, we resume transaction exits at the natural rate. 

\newtext{Our dataset does not include dust or spam transactions as Carbyne, which is not a spam filtering technique, handles both genuine and dust or spam transactions identically. It improves transaction flow and enhances the network’s capacity to handle a greater volume of transactions. Since our approach does not differentiate between benign and spam transactions, incorporating spam into the dataset would yield the same results. That said, our scheme is orthogonal to existing spam mitigation mechanisms and can complement them without altering its core functionality.}

\newtext{By demonstrating that Carbyne efficiently processes even extreme transaction volumes, we provide empirical evidence validating its DoS-resilient properties.} We consider two strategies:

\paragraph{Dimensioning for Large Transaction Volumes.}
We explore a preemptive strategy to sustain a transaction volume of 600k transactions. We consider \texttt{CbTxFilter} with expiry of sizes of 1.8\,MB$\times$2, 3\,MB$\times$2, and 12\,MB$\times$2, three times the size of filters considered in \S\ref{sec:results}. \texttt{CbTxInputsFilter} similarly expands to 1.8\,MB, 3\,MB and 12\,MB respectively. The secondary filter activates when transactions exceed 600,000.

Figs.~\ref{fig:lar1}--\ref{fig:lar3} depict how preemptive dimensioning performs. The first filter stores the first 600,000 transactions. However when the volume exceeds 600,000 the second filter is activated, marked by a pronounced decrease in the false positive rate in Figs.~\ref{fig:lar1}--\ref{fig:lar3}. The first filter expires after 14 days, and the number of transactions in the mempool returns to normal levels. Figs.~\ref{fig:lar1}--\ref{fig:lar3} shows the average false positive rate at three key points: when transactions cross 600,000, when the first filter expires, and at the end of 90 days. We see that 1.8\,MB, 3\,MB experience degradation but quickly recovers from the congestion event, however the 12\,MB filter is unperturbed throughout. As expected, the FPR is significant for the smaller filters and decreases for larger filters. The 1.8\,MB, 3\,MB and 12\,MB filters process transactions with 99.698\%, 99.914\% and 99.997\% accuracy. 

\paragraph{Dynamic Adaptation for Congestion.} 
Our next strategy considers a more dynamic strategy. We initialize a counter to track total number of transactions in \texttt{CbTxFilter}, and as soon as filter capacity of 200,000 is exceeded, additional bloom filters are generated recursively as per demand~\cite{guo2009dynamic} ~\cite{beyer2011system}. The user can tweak the filter capacity parameter as per resources available at the node. The filters can then be expired in the order of their age at defined intervals. New filters can be generated repeatedly if congestion persists.

Figs.~\ref{fig:dyn1}--\ref{fig:dyn3} show the results for filter sizes 600\,kB, 1\,MB and 4\,MB filter sizes. In all three case, additional filters are generated at the 200k, 400k, and 600k transaction marks. This means that four filters are functional at the peak of the congestion event (consuming 2.4\,MB, 4\,MB and 16\,MB respectively). Each filter expires 14 days after it was generated. When transaction volume normalizes, additional filters are no longer needed. We also highlight the average FPR at three key points: when transactions cross 600,000, when the first filter expires, and at the end of 90 days.

As expected, the false positive rate decreases with increasing filter sizes for the duration of the congestion period and otherwise: the 600\,kB, 1\,MB and 4\,MB filters process transactions with 99.222\%, 99.766\% and 99.994\% accuracy over the 90 day period. We see that the 600\,kB and 1\,MB filters experience significant performance degradation but quickly recover from the congestion event, whereas the 4\,MB filter does not deviate much. Even in the worst case scenario, for the 600\,kB filter false positive rates over the 55 hour period of congestion, the cumulative false positive rate is 4.6\E{-2}. Medium size filters of size 1\,MB can process this load with a false positive rate of approximately 1.1\E{-3}.

Both approaches perform well in combating increased spam events and have their pros and cons. The first strategy of preempting congestion performs slightly better in terms of false positives but has a larger memory footprint. The second approach dynamically adjusts to congestion and has a considerably smaller memory footprint. Overall, though a transaction load of 600,000, over three times the historic maximum witnessed on the network, is estimated to push mempool space over the 1\,GB mark, but Carbyne processes it within a few tens of MBs with extremely high accuracy.

\section{Conclusion and Future Work}
\label{sec:conclusion}
In this paper, we propose Carbyne, an ultra-lightweight counting bloom filter-based scheme to improve the performance and resilience of the Bitcoin network to increased transaction flows, network congestion and spam. 

Carbyne reduces the footprint of the mempool by two orders of magnitude while preserving key functions for processing and forwarding transactions. We devise a mechanism to expire transactions based on age, tracking and limiting double-spends, and explore potential strategies to enable Replace-by-Fee transactions. We also demonstrate strategies to cope with congestion and spam. Implementing Carbyne does not necessitate forking the network.

Carbyne supports typical real-world transaction volumes of 300\,MB in as little as 3\,MB of memory, with more than 99.9\% accuracy in processing and forwarding transactions. In simulated stress tests, Carbyne is demonstrated to cope with congestion and spam attacks with a total footprint of around 9\,MB as opposed to around 1\,GB in Bitcoin Core, and with very high fidelity. Our experimental results are obtained using Bitcoin transaction data over 90 days. This dataset is a distinct contribution, of independent interest to researchers.

In future work, we intend to further explore and fine tune trade offs in memory, accuracy, and computation. We hope to develop a functional prototype for live deployment. We are also currently adapting Carbyne to Ethereum. 

We hope this effort motivates research on the mempool and its intricacies and contributes to the security and robustness of the Bitcoin ecosystem.

\appendix

\section{\emph{MempoolState} Dataset}
\label{sec:appendix_MempoolStateDataset}

To date, data-driven research in cryptocurrencies has mostly focused on price fluctuations and economic trends, analyzing cryptocurrency transaction graphs, and mapping network topology. To the best of our knowledge there is no publicly released dataset which studies the network state of cryptocurrency networks. Recording live network state allows us to reconstruct network state at the client and replay network activity for simulation and modeling purposes. This has several useful applications which include the following: 
\emph{Mining strategies} to maximize the profit earned through transaction fee; 
\emph{Retroactively study network state} through the dataset spanning a 90-day period, with network message entries; 
\emph{Time-based snapshots} can be recreated to visualize network view for a particular node; 
\emph{Comparison among different nodes} can be made, for instance between the mempool state of different nodes; 
\emph{Identification of anomalies} can be undertaken for debugging and flagging suspicious activity; 
\emph{Forensic Investigation} can be undertaken by researchers in the wake of incidents (like the ‘dust’ attack of 2015 and onwards); 
\emph{Client-side optimizations} and security defences can be developed and tested; 
\emph{Network-Wide Impact} of community policies like pay-per-fee can be studied.

Our \emph{MempoolState} Dataset comprises three datasets, with dimensions listed in Table~\ref{tab:dataset}, and details as follows: 
\begin{itemize}
\item 
\textbf{Mempool Activity} includes all transaction entries and exits in the mempool in JSON format. Code written in C++ was used to modify Bitcoin Core, specifically \texttt{src/ txmempool.cpp}, to capture the data. For our 90-day dataset, we log $\sim$29 million unique transactions with $\sim$88 million inputs. The resulting dataset is sized at 40\,GB.

\item 
\textbf{Network Inventory} includes the \texttt{inv} messages received over the network. We log $\sim$89 million \texttt{inv} messages via the \texttt{-network flag} in the Bitcoin configuration file. The logs generated by Bitcoin included other network messages as well, such as getdata and tx. To prevent the size of the logs from becoming intractable we maintained daily log files using \emph{Logrotate} in Linux. We wrote C++ code to scan these files and extract only those \texttt{inv} messages relevant to our application. The resulting \texttt{inv} messages along with timestamp are stored in CSV format. The resulting dataset is sized at 10\,GB.

\item 
\textbf{Mempool Statistics} includes details for the raw transaction size, resulting memory usage and transaction count in the mempool. We used a Python script to invoke the Bitcoin daemon’s JSON-RPC \texttt{getmempoolinfo} method at 10 minute intervals and store the output as JSON. We use another Python script to scrape the JSON data from these files into a single CSV file for ease of plotting.
\end{itemize}

\begin{table}
\centering
\begin{tabular}{lrr}
\toprule
Content & Data Points & Size \\
\midrule
Mempool Activity & 29 million & 40 GB \\
Network Inventory & 89 million & 10 GB \\
Mempool Statistics & 12962 readings & 4 MB \\
\bottomrule
\end{tabular}
\caption{\emph{MempoolState} Dataset (Jan 01, 2021 -- 31 Mar 2021)}
\label{tab:dataset}
\end{table}

\noindent
The \emph{MempoolState} Dataset is accompanied by the following tools we have developed: 
\begin{itemize}
    \item \textbf{Parallel Simulations for Carbyne and Bitcoin} This code tool simulates the Bitcoin Core and Carbyne mempools in parallel to evaluate their performance. Carbyne simulation code is written in C++. It uses the library libbf~\cite{libbf} for counting Bloom filter and the library RapidJSON~\cite{rapidjson} to support JSON operations in C++.
    This code requires the \textbf{Mempool Activity} and \textbf{Network Activity} datasets as inputs. It logs performance at hourly intervals and generates hourly as well as cumulative stats in CSV format. It also generates forensics data to help us analyze the false positive and false negatives in detail in CSV format.
    \item \textbf{Computation Time.} This code benchmarks the querying, insertion and deletion times for \texttt{mapTx}, \texttt{mapLinks} and \texttt{mapNextTx} as well as \texttt{CbTxFilter} and \texttt{CbTxInputs Filter}. It uses C++ STL \texttt{std::chrono} to perform the computation time analysis.
    \item \textbf{Snapshot.} This code written in C++ simulates the Bitcoin Core mempool. It recreates the state of the Bitcoin Core mempool for any given time instant. It can be used for application scenarios described above.
    \item \textbf{mapRelay} This code written in C++ using the network logs recreates the mapRelay to report the average number of transactions in mapRelay over an hour.
    \item \textbf{Reconstruct} This code, written in C++ is used to check how multiple Carbyne nodes may be used to bootstrap the mempool of a newly connected node. We run experiments where Carbyne nodes store 5\%, 10\%, 20\%, 25\%, 33\% and 50\% of transactions on a random basis. Similarly, we vary the number of Carbyne nodes that a new node that joins the network connects to 4,8,12.
    \item \textbf{Carbyne.} This code, written in C++, runs the Carbyne mempool. It can be used as a starting point to prototype Carbyne or integrate it into another client. 
\end{itemize}

\section{\emph{Storing Full Transactions}}
\label{sec:appendix_optionalpool}
By optimizing the mempool, Carbyne users may still choose to store subsets of full transactions in RAM for various purposes, depending on the resources available to them.

For instance, users may store full transactions for mining. The Bitcoin Core protocol limits blocks to 1 MB in size. Carbyne nodes may maintain a live pool of 1 MB of current transactions and prioritize them on the basis of fee per size, value, age of inputs, number of ancestors/ descendants etc. and propose blocks.

A subset of transactions may be saved in RAM to bootstrap newly connecting nodes. If multiple Carbyne nodes on the network each randomly store a subset of full transactions, they can collectively bootstrap mempool of new nodes.

\begin{figure}
         \includegraphics[width=8cm]{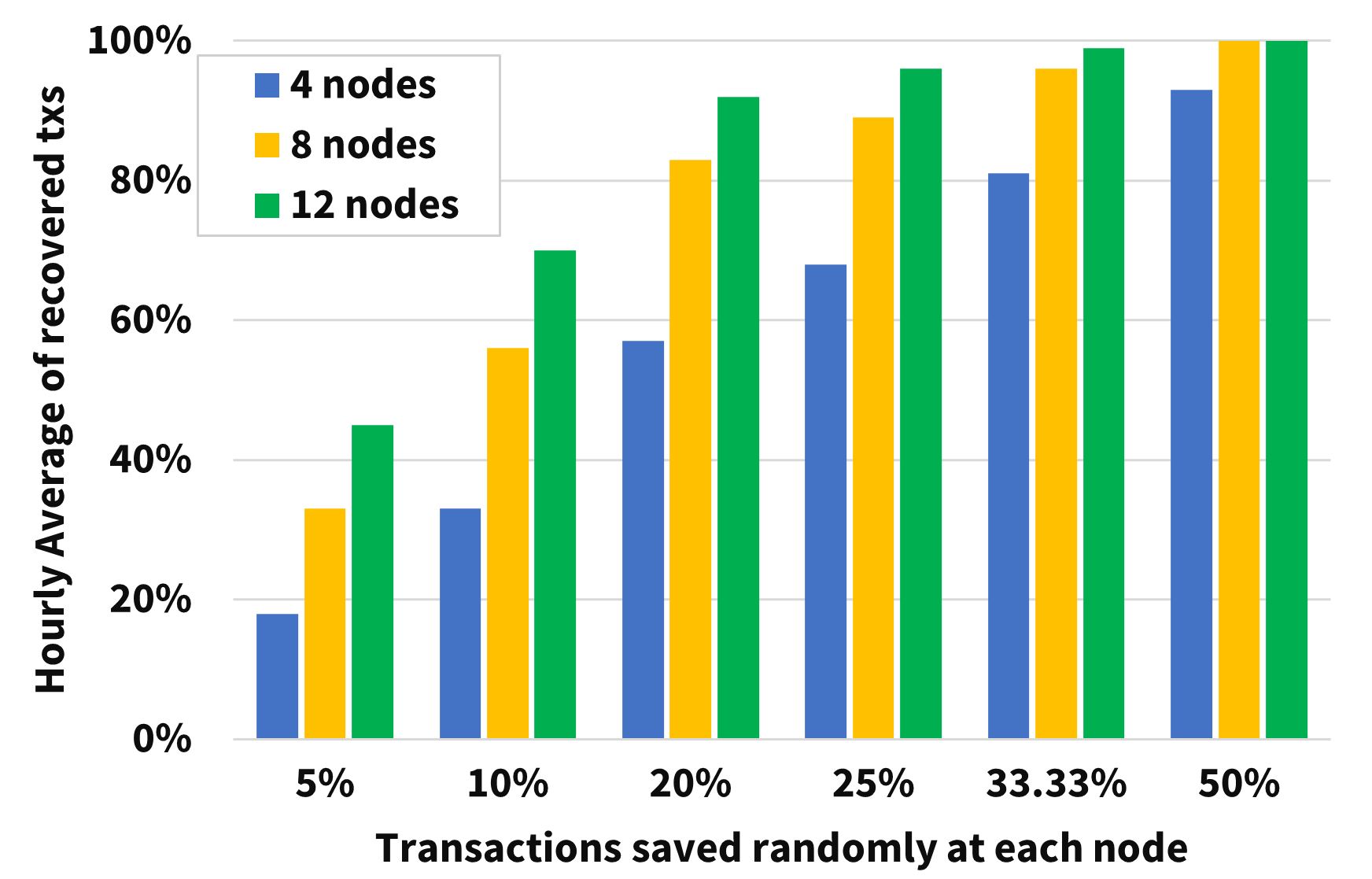}
         \caption{Recovery Rate}
         \label{fig:recover}
\end{figure}

We undertake a preliminary analysis using our \emph{MempoolState} dataset. We run experiments where Carbyne nodes store 5\%, 10\%, 20\%, 25\%, 33\% and 50\% of total transactions they receive over a 90 day period. We also vary the number of Carbyne nodes that new nodes may connect to. Our results show: if a new node connects to 4,8 and 12 Carbyne nodes each maintaining 10\% of randomly selected current transactions, the new node can recover 33\%, 56\% and 70\% mempool transactions respectively as shown in Fig.~\ref{fig:recover}.

\begin{figure}[htbp]
\centering
     \begin{subfigure}{0.48\textwidth}
         \includegraphics[width=\textwidth]{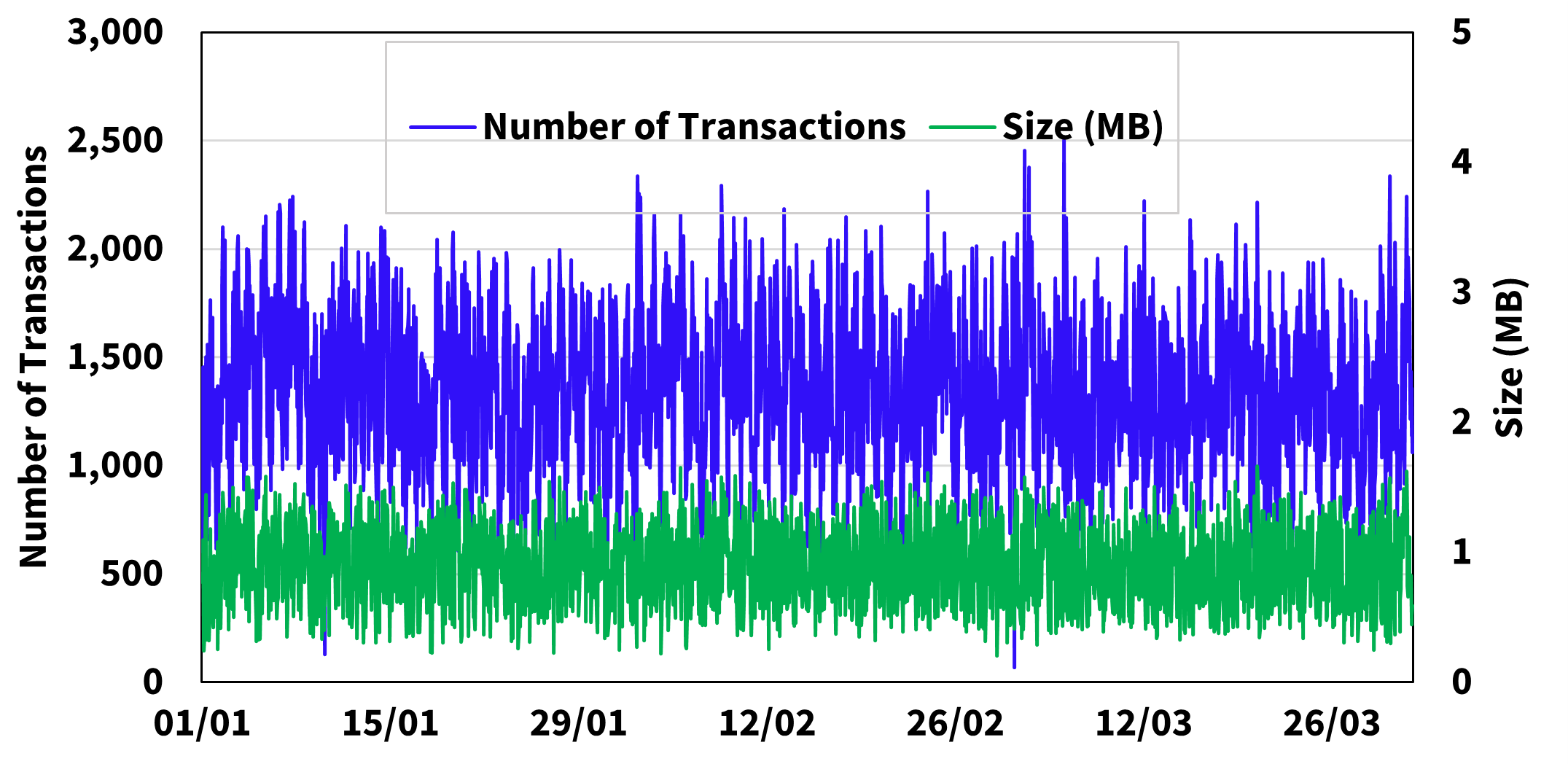}
         \caption{mapRelay (Bitcoin Core)}
         \label{fig:actual}
     \end{subfigure}
     \begin{subfigure}{0.48\textwidth}
     \vspace{3mm}
        \includegraphics[width=\textwidth]{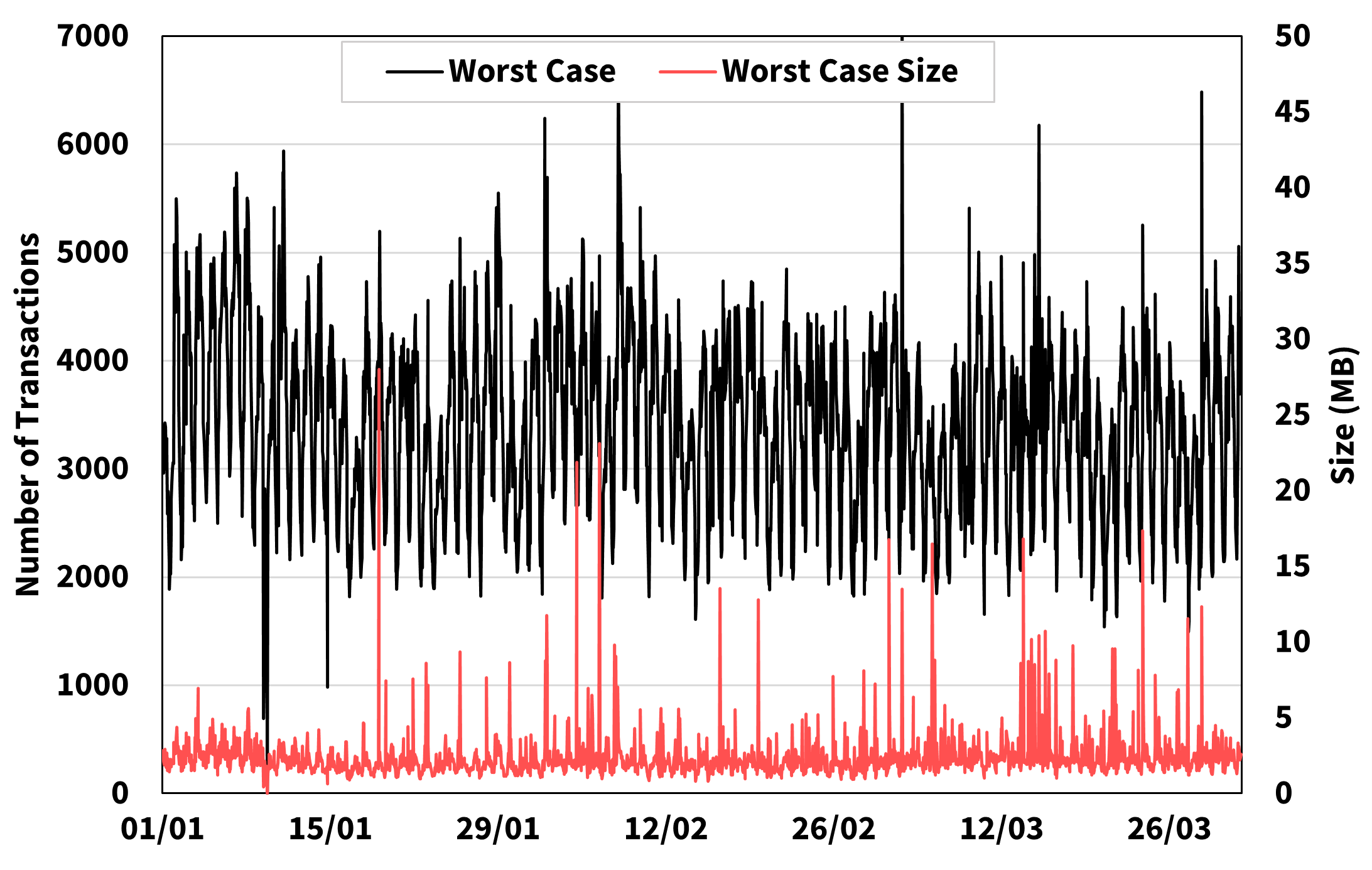}
         \caption{mapRelay (Worst Case)}
        \label{fig:worst}
     \end{subfigure}
     \hfill
      \caption{mapRelay}
       \label{fig:mapRelay}
\end{figure}

\section{\emph{Transaction Retention}}
\label{sec:appendix_txretention}
When an \texttt{entryTx} and its inputs are successfully verified, the transaction is added to Carbyne mempool as described in \S\ref{sec:proposed}. The transaction hash \texttt{TxID} is then broadcast to the node's peers with an \texttt{inv} message and full transactions are forwarded on request. In Bitcoin’s diffusion protocol (a variation on random flooding) peers inject a random delay before announcing a received transaction to its peers, to mitigate timing attacks and in-flight collisions. 

Bitcoin segregates the cache of transactions for mempool and relay in separate containers. The data structure \texttt{mapRelay} is responsible for keeping transactions until they are relayed to peers. This memory consumption comes under the umbrella of networks and connections and is in addition to the memory incurred by the mempool transactions. The length of the interval that full transactions are stored can vary : it is stored until it is relayed to all peers, or a 15 minute default expiry can also be configured by the user. This structure is part of Bitcoin Core’s network and connection handling mechanism, it is independent of the Bitcoin mempool, and our solution retains it as it is.

We undertake experiments to quantify the trade-off between retention time and memory consumption of \texttt{mapRelay}. Our node ran on default settings with 8 outbound peers and upto 125 inbound peers. Firstly, we estimate the number of transactions in \texttt{mapRelay} at any given time using the network logs collected over the 90 day period. We observe that the number of these transactions average at 1000, with a cap at around 2500 occupying no more than 2 MB at any given instant as shown in Fig. \ref{fig:actual}. Secondly, we estimate the worst case scenario where each transaction stays in \texttt{mapRelay} for 15 mins. This comes out to be 3,500 transactions on average and no more than 7000 at maximum, as shown in Fig. \ref{fig:worst}. However as empirical data shows it takes almost 13 seconds on average for a transaction to propagate 90\% Bitcoin nodes \cite{dsnresearch}. Hence we do not anticipate the memory consumption of \texttt{mapRelay} to be substantial. 

\section{\emph{Replace-by-Fee}}
\label{sec:appendix_rbf}

Replace-By-Fee (RBF) is an opt-in node policy that allows an unconfirmed transaction in a mempool to be replaced with a different transaction that spends at least one of the same inputs and pays a higher fee. RBF has multiple variants. Here we briefly discuss potential strategies for each.

\begin{itemize}
\item 
\textbf{Full RBF} unconditionally allows a transaction to replace older ones so long as it pays a sufficient fee. We can already address this within Carbyne in a sense by incrementing the threshold value of the \texttt{CbTxInputsFilter} counter to 2. The intuition here is that if a transaction input already exists in \texttt{CbTxInputsFilter}, then its respective counters will each be 1 or more. When the corresponding RBF transaction arrives, it is accepted and the counters are incremented to 2, after which more RBF transactions for this case will be dropped. A user can choose a custom threshold value. Moreover, at this point, a variable could also track the moving average of RBF transactions to detect potential misuse of the RBF provision, and halt processing these transactions.

\item 
\textbf{Opt-in RBF} allows the replacement when the transactions being replaced have explicitly signalled they allow replacement via the "sequence" field. Dual-load bloom filter could be deployed that saves the transaction inputs, along with the original fee. While Bloom filters usually hold a single type of information, which is either the membership in a given set or the return values of elements, the proposed DLBF holds both the membership and the return values in a single Bloom filter. If a node has Opt-in RBF then it can initialize two separate bloom filters for transaction inputs, one which holds inputs of replaceable transactions and the other which holds inputs of transactions that can not be replaced. A transaction should only be replaced if the inputs are in the filter that signals RBF.

\item 
\textbf{Delayed RBF} is a variant which allows transactions to be replaced unconditionally, but only after a given number of blocks have been mined since the replaced transactions were first seen by the node. A Dual-load bloom filter could be deployed that saves the transaction inputs, along with the block number when it is safe to replace the transaction.

\item 
\textbf{First-seen-safe RBF} only allows the replacement if an additional criteria is met: the replacement transaction must pay all the same outputs as the transactions being replaced. If a node employs this policy then it is best to save the complete transaction as the transaction inputs, outputs and fee all need to be stored.
\end{itemize}

\noindent We defer the detailed empirical analysis of these strategies as future work.

\end{document}